\documentclass[10pt,conference]{IEEEtran}
\IEEEoverridecommandlockouts
\usepackage[]{collab}
\usepackage{cite}
\usepackage{amsmath,amssymb,amsfonts}
\usepackage{algorithmic}
\usepackage{graphicx}
\usepackage{textcomp}
\usepackage{xcolor}
\usepackage{array}
\usepackage{textcomp}
\usepackage{float}
\usepackage{multirow}
\usepackage{tikz}
\usepackage{tcolorbox}
\usepackage{url}
\usepackage{listings}
\usepackage{threeparttable}
\usepackage{amsthm}

\usepackage{booktabs} 
\usepackage{soul}
\usepackage{ulem}
\usepackage{enumitem}
\usepackage{mdframed}
\usepackage{balance}
\usepackage{pifont}
\usepackage[font=small,bf]{caption}

\definecolor{ballblue}{rgb}{0.13, 0.67, 0.8}
\definecolor{grey}{rgb}{0.9, 0.9, 0.9}
\definecolor{googlered}{rgb}{0.914, 0.262, 0.207}
\definecolor{dandelion}{rgb}{0.95, 0.65, 0.0}
\definecolor{citecolor}{RGB}{106, 34, 107}

\usepackage[colorlinks,linkcolor=citecolor,
            anchorcolor=black,
            filecolor=black,      
            urlcolor=ballblue,
            citecolor=citecolor,
            bookmarks=false
            ]{hyperref}

\newcommand\nm{AlertGuardian\xspace}
\newcommand\company{\textit{Company-X}\xspace}

\newcommand{\ie}{{\em i.e.},\xspace}
\newcommand{\eg}{{\em e.g.},\xspace}
\definecolor{mygreen}{HTML}{AFCFA5}
\newcounter{deployment}
\newcommand{\deployment}[1]{\refstepcounter{deployment}
	\begin{mdframed}[linecolor=gray!25,roundcorner=12pt,backgroundcolor=mygreen!20,linewidth=3pt,innerleftmargin=2pt, leftmargin=0cm,rightmargin=0cm,topline=false,bottomline=false,rightline=false,leftline=false]
		\textbf{Deployment Experience \arabic{deployment}:} #1
	\end{mdframed}
	\vspace{-1mm}
}

\begin{document}

\title{AlertGuardian: Intelligent Alert Life-Cycle Management for Large-scale Cloud Systems}

\author{\IEEEauthorblockN{Guangba Yu}
\IEEEauthorblockA{\textit{Sun Yat-Sen University} \\
Guangzhou, China \\
yugb5@mail2.sysu.edu.cn}
\and
\IEEEauthorblockN{Genting Mai}
\IEEEauthorblockA{\textit{Sun Yat-Sen University} \\
Guangzhou, China \\
maigt3@mail2.sysu.edu.cn}
\and
\IEEEauthorblockN{Rui Wang}
\IEEEauthorblockA{\textit{Tencent} \\
Shenzhen, China \\
amurorywang@tencent.com}
\and
\IEEEauthorblockN{Ruipeng Li}
\IEEEauthorblockA{\textit{Tencent} \\
Shenzhen, China \\
tristonli@tencent.com}
\and
\IEEEauthorblockN{Pengfei Chen}
\IEEEauthorblockA{\textit{Sun Yat-Sen University} \\
Guangzhou, China \\
chenpf7@mail.sysu.edu.cn}
\and
\IEEEauthorblockN{Long Pan}
\IEEEauthorblockA{\textit{Tencent} \\
Shenzhen, China \\
hydrapan@tencent.com}
\and
\IEEEauthorblockN{Ruijie Xu}
\IEEEauthorblockA{\textit{Tencent} \\
Shenzhen, China \\
rjxu@tencent.com}
}

\author{\large Guangba Yu\IEEEauthorrefmark{2}, Genting Mai\IEEEauthorrefmark{2}, Rui Wang\IEEEauthorrefmark{3}, Ruipeng Li\IEEEauthorrefmark{3}, Pengfei Chen\IEEEauthorrefmark{2}$^{*}$\thanks{\hspace{-2ex}$^{*}$Corresponding author (chenpf7@mail.sysu.edu.cn).}, Long Pan\IEEEauthorrefmark{3}, Ruijie Xu\IEEEauthorrefmark{3}\\
\IEEEauthorblockA{
\IEEEauthorrefmark{2} Sun Yat-sen University, Guangzhou, China
\IEEEauthorrefmark{3} Tencent, Shenzhen, China\\
}
}

\maketitle

\begin{abstract}
Alerts are critical for detecting anomalies in large-scale cloud systems, ensuring reliability and user experience. However, current systems generate overwhelming volumes of alerts, degrading operational efficiency due to ineffective alert life-cycle management. This paper details the efforts of \company to optimize alert life-cycle management, addressing alert fatigue in cloud systems. We propose \nm, a framework collaborating large language models (LLMs) and lightweight graph models to optimize the alert life-cycle through three phases: \textit{Alert Denoise} uses graph learning model with virtual noise to filter noise, \textit{Alert Summary} employs Retrieval Augmented Generation (RAG) with LLMs to create actionable summary, and \textit{Alert Rule Refinement} leverages multi-agent iterative feedbacks to improve alert rule quality. Evaluated on four real-world datasets from \textit{Company-X}'s services, \nm significantly mitigates alert fatigue (94.8\% alert reduction ratios) and accelerates fault diagnosis (90.5\% diagnosis accuracy). Moreover, \nm improves 1,174 alert rules, with 375 accepted by SREs (32\% acceptance rate). Finally, we share success stories and lessons learned about alert life-cycle management after the deployment of \nm in \company.
\end{abstract}


\section{Introduction}


Cloud systems, such as Microsoft Azure, Amazon AWS, and Tencent Cloud, provide critical services to users worldwide. Ensuring their reliability is paramount to prevent user dissatisfaction and revenue loss. However, failures, such as unexpected interruptions or service level degradation, are inevitable in complex cloud environments~\cite{li2022incident,socc2022incident,changerca2024fse,yu2020microscaler2}. To detect and address these failures promptly, modern cloud systems employ comprehensive monitoring mechanisms that continuously track the health of services, generating diverse monitoring data (\eg key performance indicators (KPIs))~\cite{nezha2023fse,logreducer,MicroSketch,yu2021microrank}. Site Reliability Engineers (SREs) leverage their expertise and historical monitoring data to configure alert rules~\cite{chendynamic,azurerule,gu2025argosagentictimeseriesanomaly,prometheus}. When monitoring data meets these rule conditions, alerts are triggered to notify SREs for inspection and mitigation.

\begin{figure}
    \centering
    \includegraphics[width=1\linewidth]{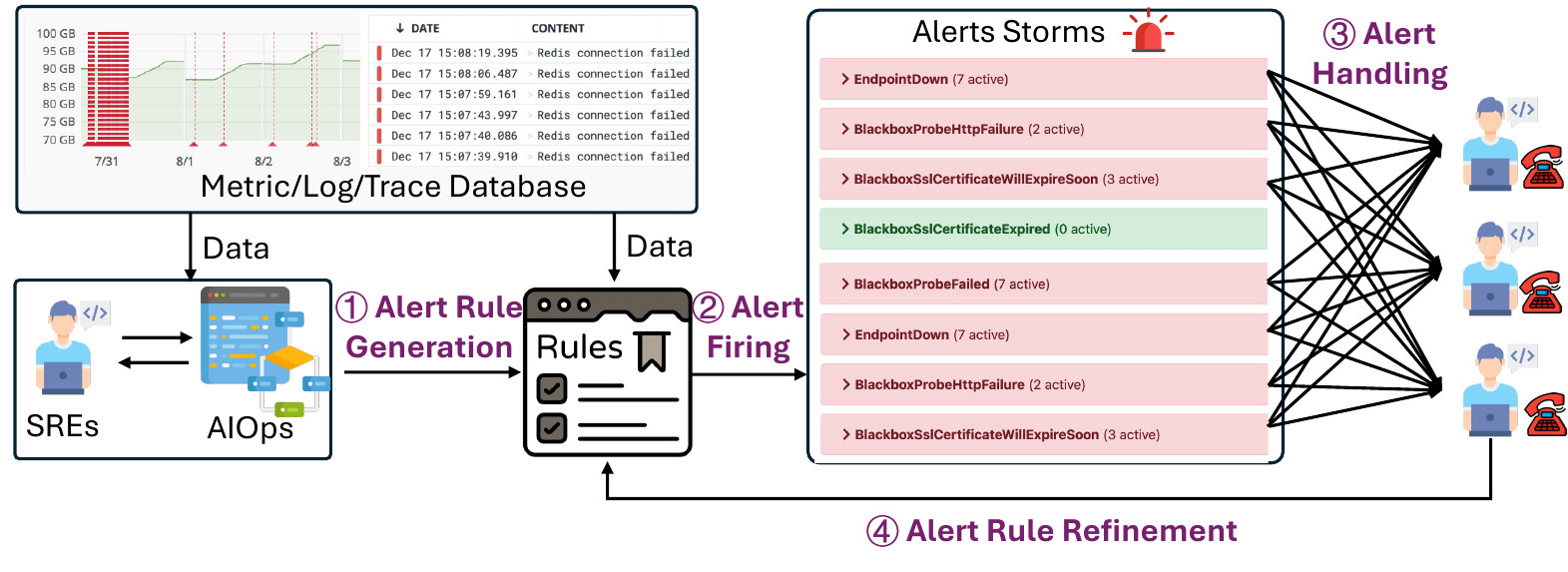}
    \vspace{-0.3in}
    \caption{The process of alert life-cycle management.} 
    \vspace{-0.3in}
    \label{fig:problem}
\end{figure}

Although alert systems have evolved significantly from their humble beginnings, they continue to present substantial challenges that hinder operational effectiveness across organizations~\cite{man2012alert,xu2017lightweight,chen2022online,mahdavi2020real,zhao2020understanding,He2022GIED}. As cloud systems grow more complex and interconnected, the sheer volume of alerts generated by monitoring systems has reached unprecedented levels. Modern monitoring infrastructures generate thousands of alerts daily, creating an overwhelming deluge of notifications that often obscure critical issues rather than illuminating them. This phenomenon, commonly referred to as ``alert fatigue''~\cite{aminanto2019automated}, has become a significant impediment to operational efficiency across industries. The fundamental problem of alert fatigue lies in poor alert life-cycle management (\S~\ref{sec:problem}). As shown in Fig.~\ref{fig:problem}, the alert life-cycle encompasses \textit{Alert Rule Generation}, \textit{Alert Firing},  \textit{Alert Handling} and \textit{Alert Rule Refinement}. Existing alert systems, such as Prometheus Alertmanager~\cite{alertmanager}, focus primarily on alert rule configuration and alert firing, neglecting the quality of rules and alerts. This leads to poorly managed rules and frequent alert storms.

Recent efforts in alert denoising~\cite{turgeman2022context,zhao2020automatically, zhao2020understanding, chen2021graph} attempt to reduce SRE burden by identifying noise and correlating alerts during the Alert Handling stage. However, with the proliferation of alert attribute combinations, traditional denoising methods struggle to adapt to modern alert complexity (\S~\ref{limit1}). Even after denoising, alerts typically flag deviations without providing the essential context needed to understand their implication or root causes. This contextual vacuum forces SREs to engage in time-consuming investigations, combining fragmented information from disparate sources to construct a coherent understanding of the underlying problem (\S~\ref{limit2}). Moreover, real-world systems also experience dynamic changes in traffic patterns and operating environments, making static rules prone to persistent false alarms. Optimizing rules manually requires extensive expert knowledge and is labor-intensive, underscoring the need for intelligent automated solutions (\S~\ref{limit3}).

\textbf{AlertGuardian.} To address the limitations of existing alert systems, this paper presents the alert life-cycle management practices at \company, a leading Internet service provider with hundreds of millions of users. We introduce \nm, a framework collaborating large language models (LLMs) and lightweight graph models to optimize the alert life-cycle. To meet real-time analysis demands, \nm first leverages graph learning model with the virtual
noise to denoise alerts, accommodating variable attributes and filtering noise effectively (\S~\ref{sec:denoise}). It then integrates Retrieval-Augmented Generation (RAG) with LLMs based on internal knowledge (\eg system documents, alert rule explanations, incident tickets) to transform cryptic alerts into comprehensive narratives that provide both what and why of emerging issues (\S~\ref{sec:summary}). Additionally, to thoroughly address alert noise, an offline multi-agent workflow with iterative feedback optimizes rules through deduplication, threshold adjustments, and temporal analysis (\S~\ref{sec:refine}).

\textbf{Results.} We conduct a comprehensive study on four real-world datasets from representative services (gaming, office, media, and education) at \company (\S~\ref{sec:exp}). Experimental results show that \nm achieves high alert reduction ratios (93.82\% to 95.50\%) across all systems, significantly outperforming baseline methods. Moreover, \nm provides high-quality alert summaries (98.5\% action accuracy) that reduce failure diagnosis overhead. \nm also improves 1,174 alert rules, with 375 accepted by SREs (32\% acceptance rate). 
The denoise module has been deployed at \company for more than one year, with summary and rule refinement modules in pilot use for over three months. The deployment of \nm offers practical insight for addressing alert fatigue in large-scale cloud systems.

\textbf{Contribution.} This paper makes following contributions.
\begin{itemize}[leftmargin=*]
    \item We propose addressing alert storm from an alert life-cycle perspective, and share practical experiences from \company, providing valuable insights for future research and implementations in alert systems.
    \item We propose \nm, a novel framework that collaborates LLMs and lightweight graph models to optimize the alert life-cycle management in practice.
    \item We evaluate \nm on four real-world datasets from \company, demonstrating robust performance and practical impact across industrial workloads. 
\end{itemize}


\section{Background and Motivation}

\subsection{Background}\label{sec:background}
Alert mechanisms are fundamental to monitoring systems, enabling timely anomaly detection in cloud systems. We first introduce the background about alert management. 

\begin{figure}
	\centering
	\includegraphics[width=1\linewidth]{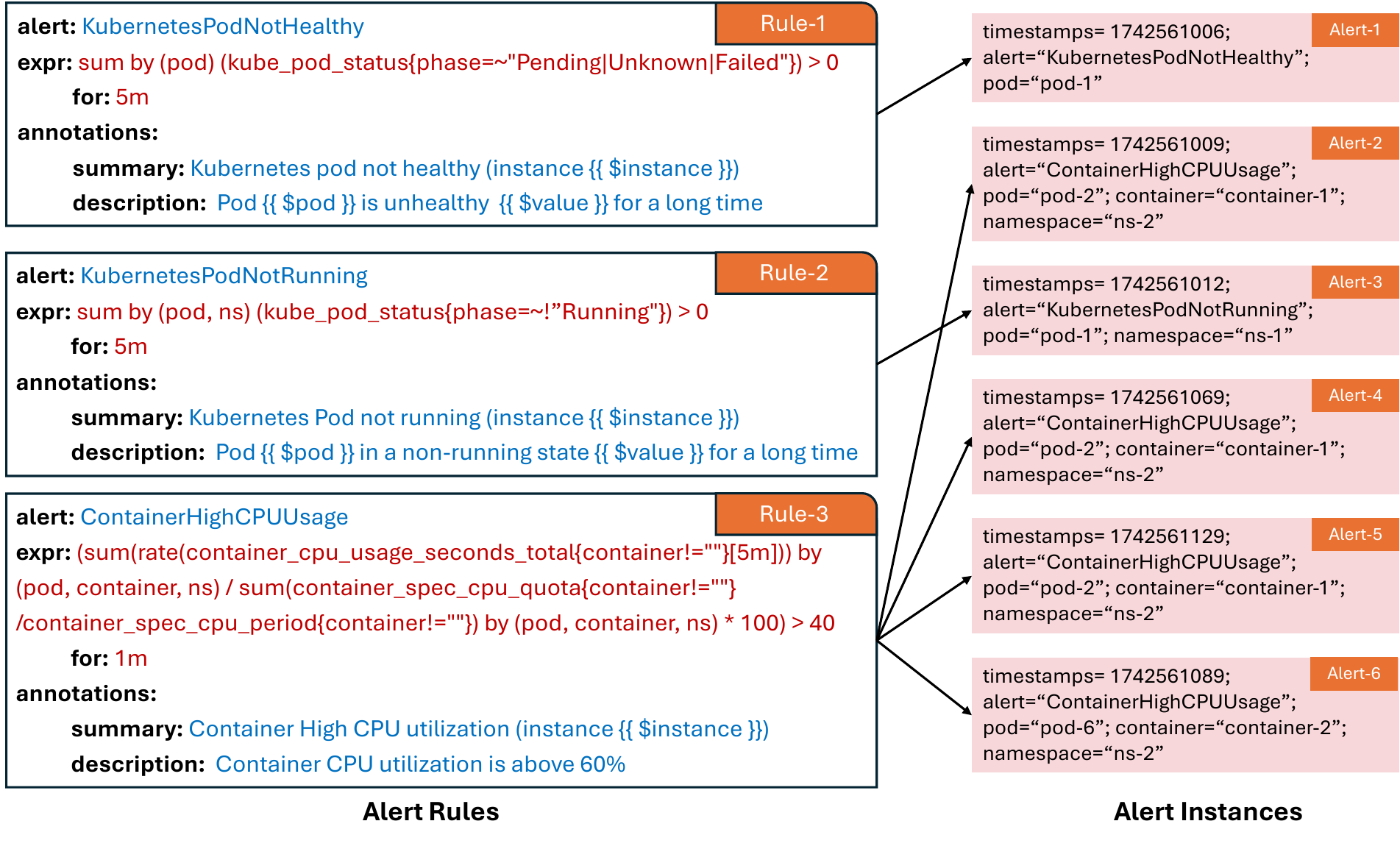}
    \vspace{-0.25in}
    \caption{Examples of alert rules and alert instances.} 
    \vspace{-0.3in}
    \label{fig:alertrule}
\end{figure}

\textbf{Alert Rules} define conditions under which notifications are triggered, typically based on expressions written in a query language such as Prometheus Query Language (PromQL)~\cite{PromQL} and LogsQL~\cite{LogsQL}. These rules periodically evaluate a specified query against a data source from monitor databases (\eg Promethuse~\cite{prometheus}), retrieving monitor data for analysis. As examples in Fig.~\ref{fig:alertrule}, an alert rule comprises four core components: (1) a query that selects the dataset to evaluate, with syntax dependent on the data source (\eg PromQL for Prometheus); (2) a condition, such as a threshold, that must be satisfied to activate the alert (\eg CPU Utilization $> 80\%$); (3) an evaluation interval and duration, determining how frequently the rule is checked (\eg 1 minute) and how long the condition must persist to trigger an alert instance; (4) some annotations provide extra details for alert responders to help them understand potential issues.

\textbf{Alert Instances} (hereafter referred to simply as ``alert'') are fired when the condition of alert rules is met.
Each alert is identified by a unique set of attribute pairs, enabling the system to differentiate among multiple alerts triggered by the same rule. As examples in Fig.~\ref{fig:alertrule}, an alert rule monitoring pod CPU usage could produce alerts with attribute sets \{\texttt{alert="PodHighCPUUsage", pod="pod-1"}\} and \{\texttt{alert="PodHighCPUUsage", pod="pod-2"}\}, representing distinct events due to differing \texttt{pod} attribute. The attribute facilitates searching, silencing, and routing of notifications, ensuring precise alert management. Furthermore, annotations, which are another set of key-value pairs, provide contextual details (\eg alert explanation) to aid responders in diagnosing and resolving issues. 

\begin{figure*}
  \begin{minipage}[t]{0.24\textwidth}
    \centering
    \raisebox{\dimexpr-\height+\ht\strutbox}{\includegraphics[width=1\textwidth]{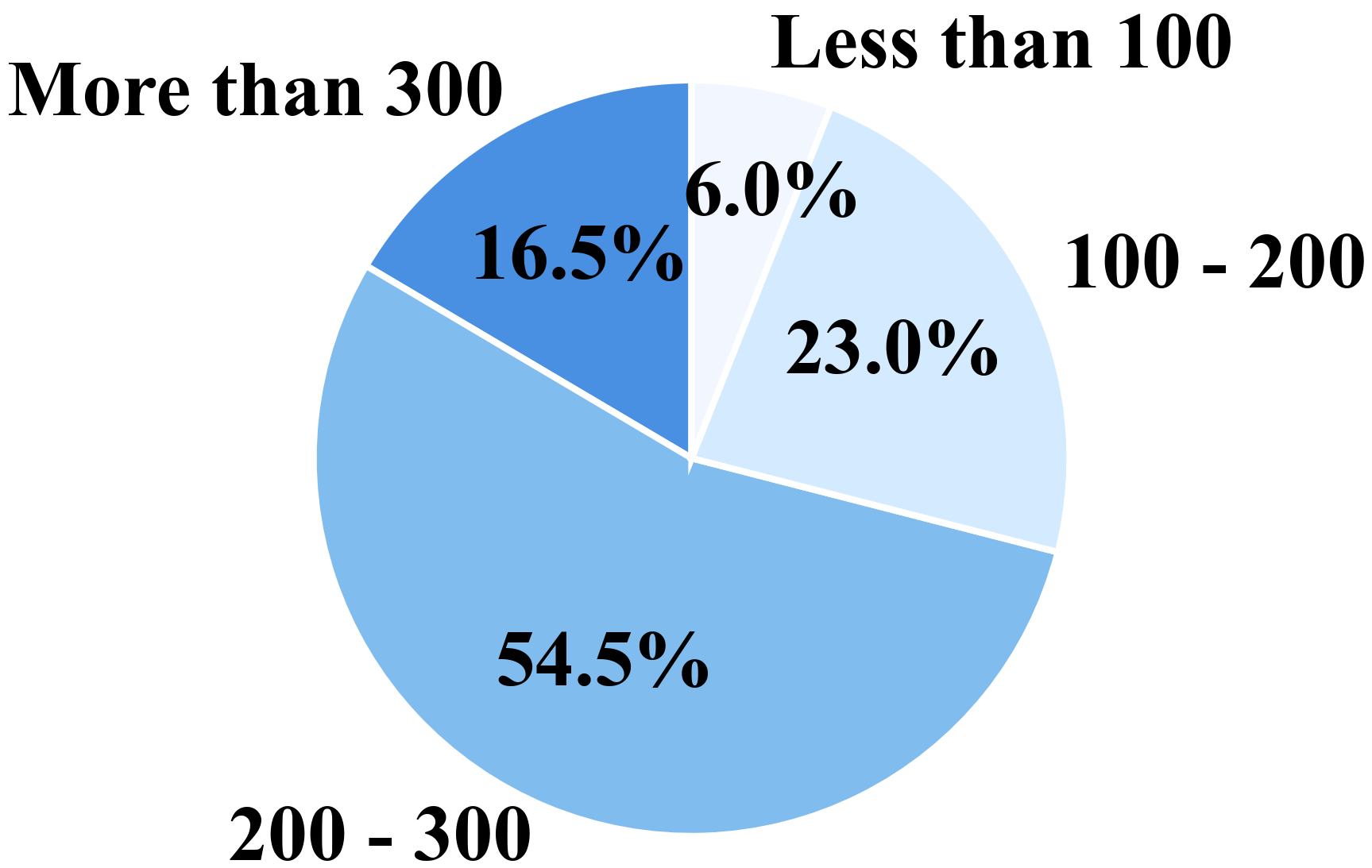}}
	\vspace{-0.08in}
	\caption{The percentage breakdown of average per-minute alert volume in \company.}
 	\label{fig:pie_alert}
  \end{minipage}
  \quad
  \begin{minipage}[t]{0.34\textwidth}
    \centering
    \raisebox{\dimexpr-\height+\ht\strutbox}{\includegraphics[width=1\textwidth]{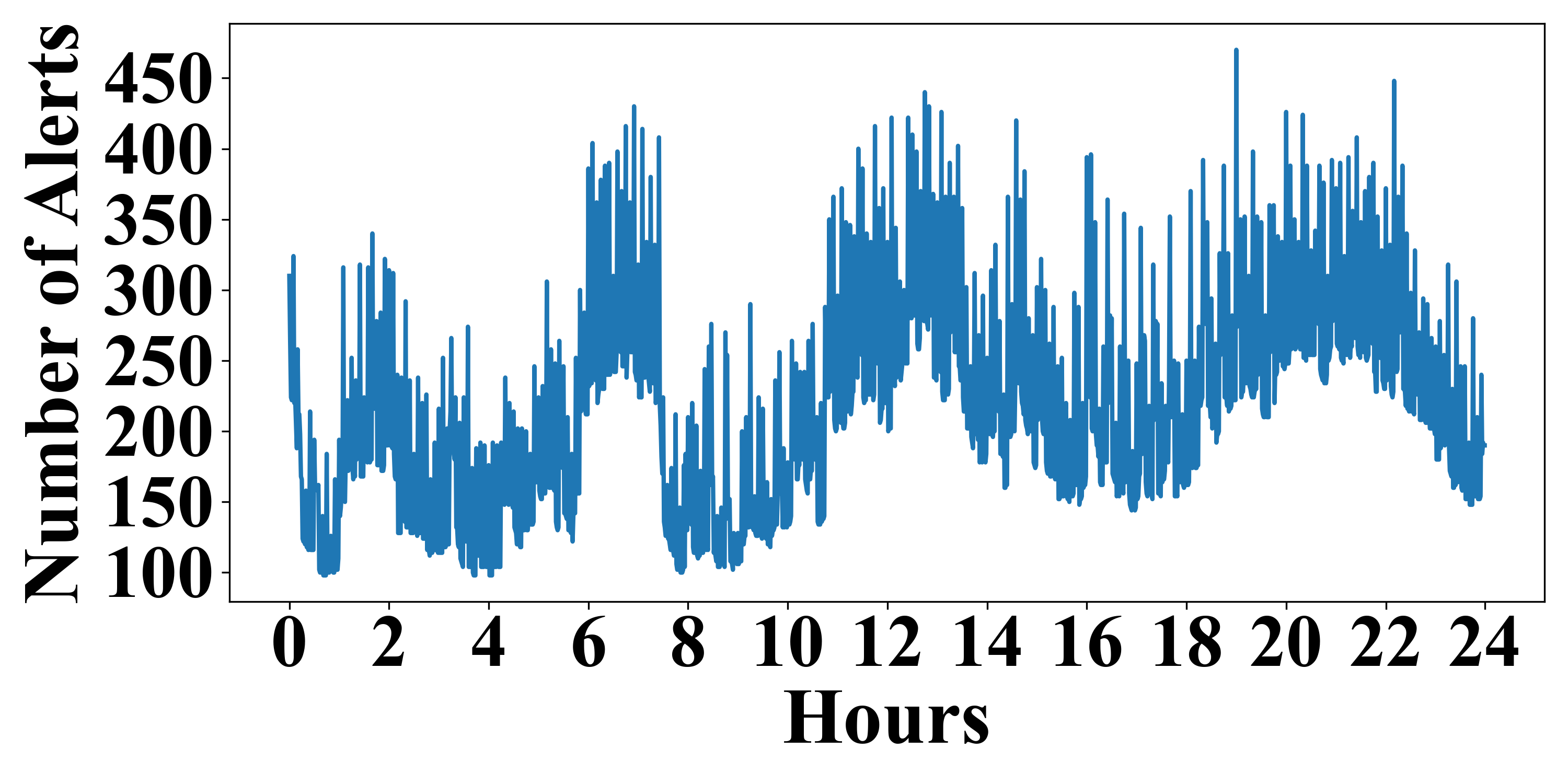}}
	\vspace{-0.08in}
	\caption{Change trend of alert volume under 59,607 alert rules in system \textit{A}.}
 	\label{fig:per_day_alert}
  \end{minipage}
  \quad
  \begin{minipage}[t]{0.34\textwidth}
    \centering
    \raisebox{\dimexpr-\height+\ht\strutbox}{\includegraphics[width=1\textwidth]{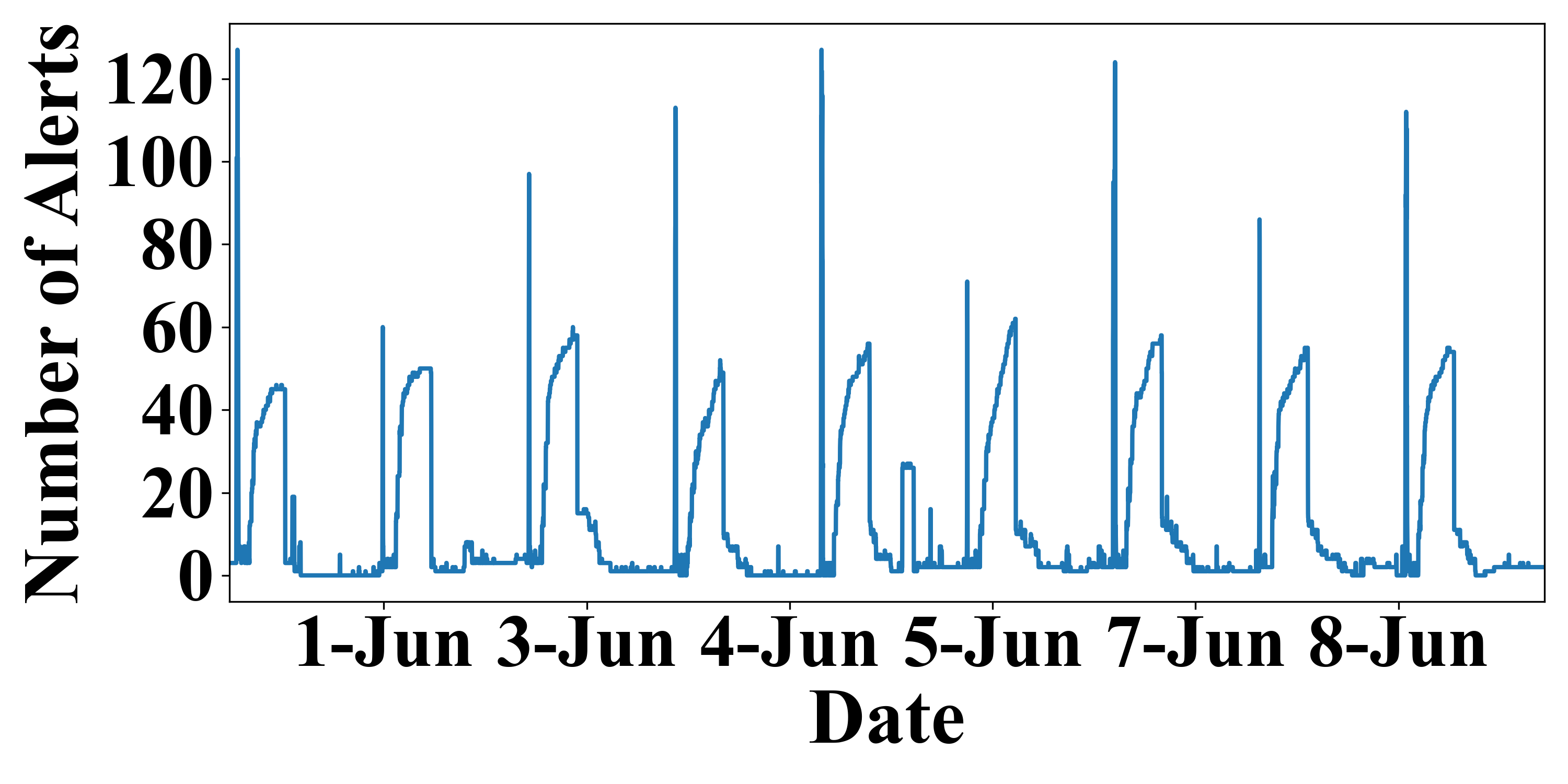}}
    \vspace{-0.08in}
	\caption{Change trend of alert volume under 3,544 alert rules in system \textit{B}.}
 	\label{fig:periodicity}
  \end{minipage}
  \vspace{-0.25in}
\end{figure*}

    

\textbf{Alert Life-Cycle Management} covers the entire process from alert rule creation to alert firing, handling, and feedback, ensuring continuous monitoring and optimization of alerts.
As illustrated in Fig.~\ref{fig:problem}, alert management generally follows a cyclical progression comprising four stages:
\begin{enumerate}[leftmargin=*, label={\ding{\numexpr171+\arabic{enumi}\relax}}]
\item \textbf{Alert Rule Generation.}
In this phase, alert rules are established to monitor various data sources, including metrics, logs, and traces stored in monitoring databases. These rules are often configured by SREs to detect potential failures. 
\item \textbf{Alert Firing.}  
Once the monitor data meet designated rules, the alerting system fires the corresponding alerts. This step forms a logical bridge between passive monitoring and active response, converting raw measurements into actionable signals. The design of thresholds and conditions here is crucial, as it directly influences both the volume and the fidelity of triggered alerts.
\item \textbf{Alert Handling.}  
Following the firing of alerts, they are routed to SRE teams or automated workflows for thorough investigation. During this stage, SREs review the alerts to determine root causes, assess severity, and execute remediation plans. Detailed logging of all diagnostic and corrective actions is essential for knowledge retention, collaboration, and post-incident analysis. By efficiently triaging and addressing alerts, organizations can minimize downtime and optimize resource allocation.
\item \textbf{Alert Rule Refinement.}  
Insights gleaned from the alert handling process inform the continuous refinement of alert rules. By analyzing patterns of false positives, false negatives, and duplicate notifications, teams can adjust thresholds or query parameters. This iterative practice reduces alert fatigue by eliminating noisy alerts while retaining sufficient sensitivity to capture anomalies.
\end{enumerate}

\subsection{Problems in Our Existing Alert Management System}\label{sec:problem}

Our institution, a large-scale Internet company (hereafter referred to as \company), maintains extensive cloud infrastructures to support a variety of globally popular products with hundreds of millions of users. These products span multiple domains, including gaming, office productivity, and education.
In an effort to manage the alerts generated by more than 200 distinct systems, our SRE team at \company built an in-house alert system. This platform facilitates the creation of alert rules, firing of alerts, and routing of notifications to relevant teams, thereby alleviating some of the operational load on SREs. However, as our systems continue to grow in both size and complexity, several significant limitations have surfaced in the existing alert management framework.

To better understand these issues, we collected nine days worth of alert data and all alert rules from more than 200 systems in \company, producing a dataset that exceeds 20 GB of alert and 200,000 alert rules in total. In the following, we show the key problems revealed by this investigation.

\subsubsection{\textbf{Problem 1: Frequent Alert Storms}}\label{problem1}
According to the experience of our SRE team, they are inundated daily by alert storms that generate an overwhelming volume of notifications in short timeframes. This phenomenon not only increases the risk of missing critical alerts due to cognitive overload, but also exacerbates the potential for alert fatigue, wherein genuinely critical signals can be overlooked amid a deluge of non-critical ones. As shown in Fig.~\ref{fig:pie_alert}, more than 70\% systems fire at least 200 alerts per minute on average, resulting in daily alert counts that can exceed 288,000 per system. Such high-frequency alerts strain SREs and delay diagnosis and remediation.

To better determine which alerts warrant reduction, we correlated system incident records with detailed alert data. The results reveal a marked discrepancy between the number of actual failures and total alerts: some systems generate tens of thousands of daily alerts without experiencing any failures. Further investigation indicates that many of these alerts represent mere ``noise'', triggered independently of failures. They can be classified into two main categories:
\begin{itemize}[leftmargin=*]
    \item \textbf{Persistent Alerts} remain active whether or not the system is functioning normally. As shown in Fig.~\ref{fig:per_day_alert}, a high volume of these alerts recurs daily without any concrete anomalies in the system. Typically, they arise from overly sensitive thresholds in alert rules (\eg a threshold is set so low that normal fluctuations are flagged as anomalies), resulting in continuous false positives. 
    \item \textbf{Periodic Alerts} emerge at regular intervals, leading to recurring alert storms. For example, as depicted in Fig.~\ref{fig:periodicity}, scheduled night-time restarts on a particular host cause a surge in alerts each day. Because these restarts are not indicative of failures, these alerts similarly constitute noise. 
\end{itemize}
Such noisy alerts consume SRE resources during the Alert Handling phase, hampering effective diagnosis. Consequently, a dedicated denoising step is essential to filter out irrelevant alerts while retaining those associated with system failures.


\subsubsection{\textbf{Problem 2: Low-quality Alert Rules}}\label{problem2}
To investigate the root causes behind frequent alert storms, we analyzed the alert rules responsible for triggering these alerts. Our findings reveal that the system contains an overwhelming number of alert rules. For example,  as shown in Fig.~\ref{fig:per_day_alert}, system \textit{A} alone has 59,607 active alert rules.
A primary contributor to excessive alert noise lies in the low quality of these rules, which are manually created and suffer from two major deficiencies:
\begin{itemize}[leftmargin=*]
  \item \textbf{Functionally Redundant Rules.}  
  Due to the lack of standardized rule design guidelines, multiple alert rules within \company often overlap in functionality. For instance, \texttt{Rule-1} and \texttt{Rule-2} in Fig.~\ref{fig:alertrule} both aim to detect whether a \texttt{pod} is not in the \emph{Running} state, yet they are implemented separately and generate distinct alerts. In addition, different engineers can define rules using varying combinations of attributes (\eg \texttt{pod}, \texttt{namespace}, \texttt{container}), resulting in duplicate alerts that differ only in attribute values.
  \item \textbf{Inappropriate Static Thresholds.} Alert rules in \company frequently rely on fixed thresholds (\eg a CPU utilization threshold of 40\% as shown in Fig.~\ref{fig:alertrule}), which do not adapt to dynamic system behaviors or workload variations. As a result, these rigid thresholds often lead to false positives (\ie normal fluctuations being misclassified as anomalies) which ultimately flood SREs with irrelevant alerts.
\end{itemize}
These issues highlight the urgent need for a \emph{Alert Rule Refinement} process to eliminate redundancy, adapt thresholds to evolving environments, and reduce unnecessary alert noise. 


\begin{figure*}
  \begin{minipage}[t]{0.31\textwidth}
    \centering
    \raisebox{\dimexpr-\height+\ht\strutbox}{\includegraphics[width=1\textwidth]{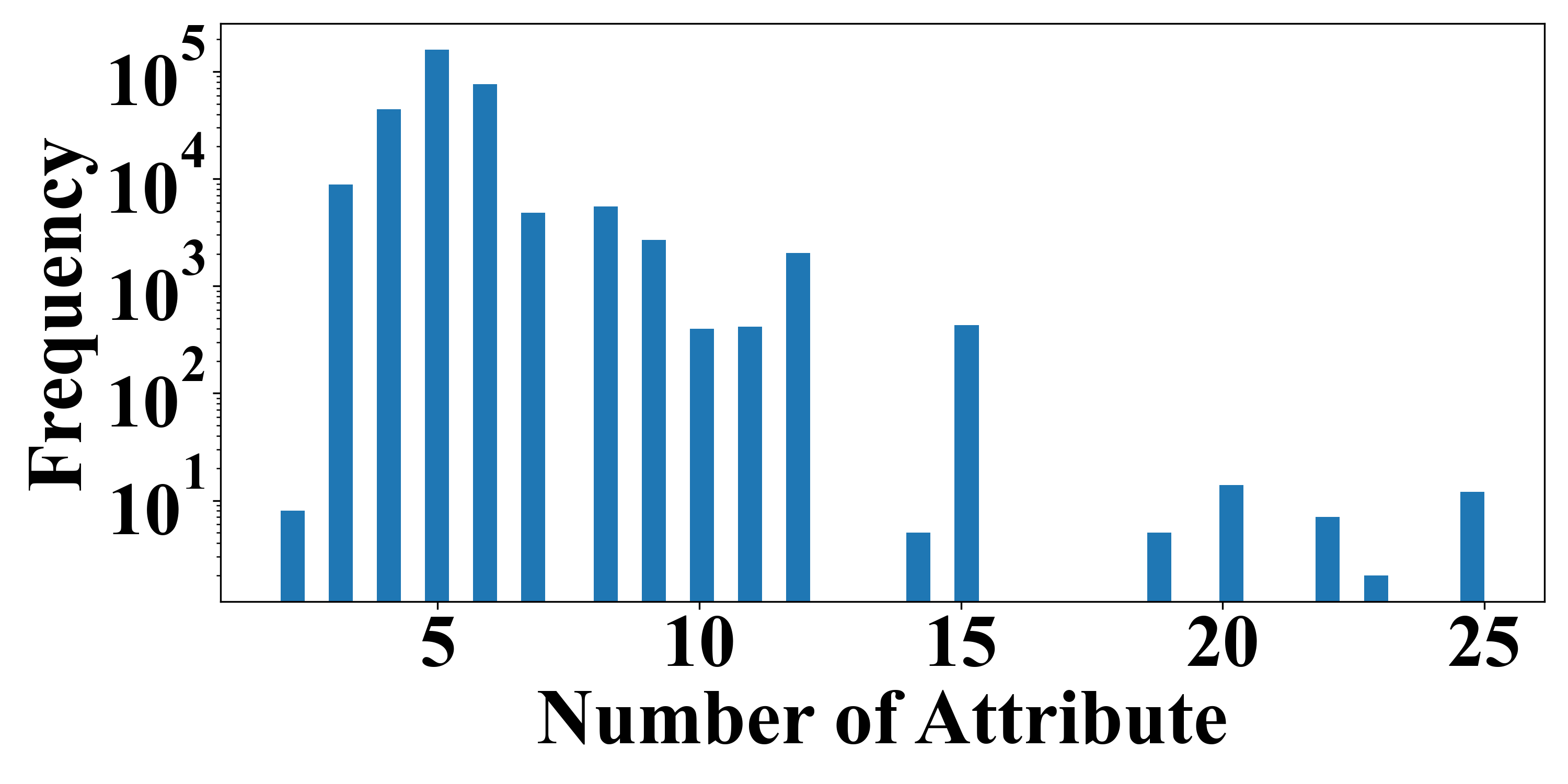}}
	\vspace{-0.08in}
	\caption{Distribution of the number of alert attributes in \company.}
 	\label{fig:attribute_num_distribution}
  \end{minipage}
  \quad
  \begin{minipage}[t]{0.31\textwidth}
    \centering
    \raisebox{\dimexpr-\height+\ht\strutbox}{\includegraphics[width=1\textwidth]{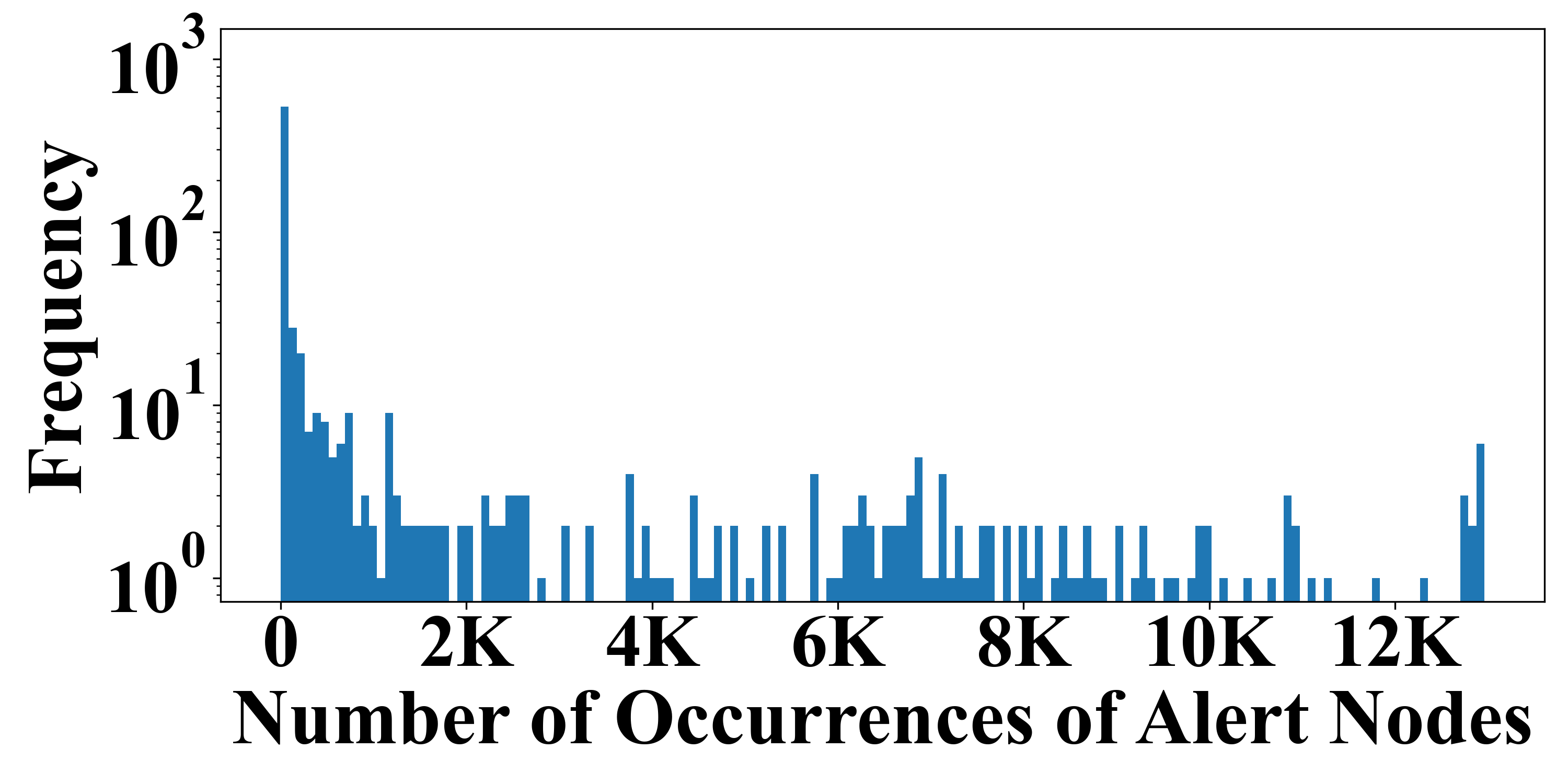}}
	\vspace{-0.08in}
	\caption{Distribution of occurrences for alert monitoring nodes in system \textit{C}.}
 	\label{fig:occurrence}
  \end{minipage}
  \quad
  \begin{minipage}[t]{0.31\textwidth}
    \centering
    \raisebox{\dimexpr-\height+\ht\strutbox}{\includegraphics[width=1\textwidth]{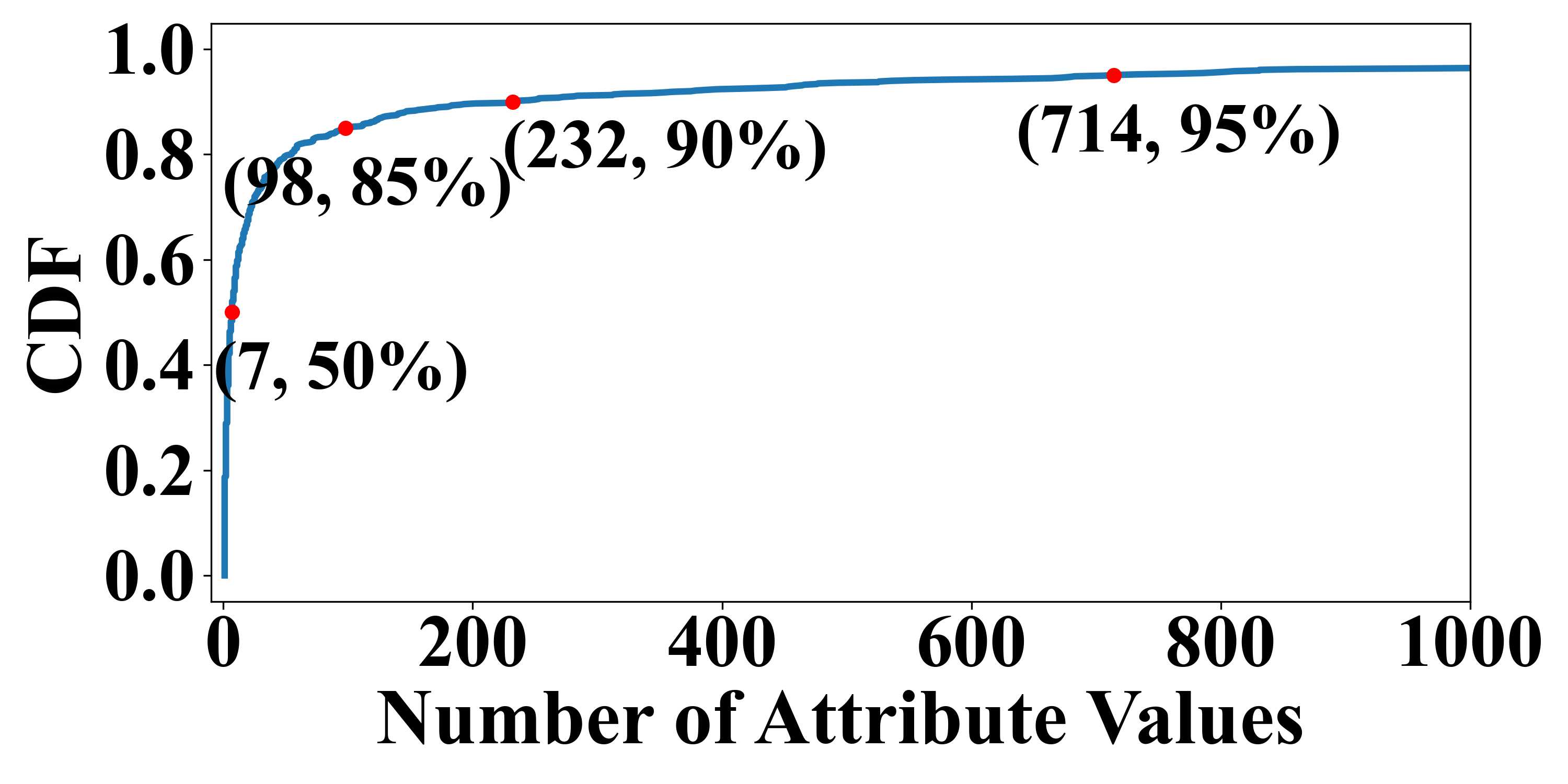}}
    \vspace{-0.08in}
	\caption{CDF of the number of  values of all alert attributes in \company.}
 	\label{fig:cdf}
  \end{minipage}
  \vspace{-0.25in}
\end{figure*}

\subsection{Why Our Existing Alert Systems Do Not Help?}\label{sec:existing}

\subsubsection{\textbf{Limitation 1: Obsolete Alert Denoising Mechanism}}\label{limit1}
In our existing Alert Management System, the alert denoising module predominantly handles alerts defined by fixed attribute combinations~\cite{turgeman2022context, zhao2020understanding, chen2021graph}. The primary denoising strategy relies on pairwise co-occurrence frequencies among monitored entities, which was adequate for earlier simplistic systems. However, as the number of managed systems increases and their underlying architectures become more complex, alerts in \company now encompass significantly broader and more diverse attributes (see Fig.~\ref{fig:attribute_num_distribution}). Consequently, the proliferation of attribute combinations exacerbates the issue of low pairwise co-occurrence frequencies, with many alert entities co-occurring only a few times (see Fig.~\ref{fig:occurrence}). In addition to the high volume of attribute combinations, certain attributes, such as IP addresses and Pod IDs, have extensive value ranges. As illustrated in Fig.~\ref{fig:cdf}, over 90\% of attributes possess more than 200 potential values, exceeding the capability of the current denoising approach to handle such large-scale variability. 
As illustrated in Fig.~\ref{fig:cdf}, over 90\% of attributes possess more than 200 potential values, which exceeds the current denoising approach's capability to handle such large-scale variability, leading to Problem 1 (§\ref{problem1}). Therefore, a more adaptable alert denoising method is imperative, one capable of accommodating the vast number of attributes and their increasingly diverse value distributions.

\subsubsection{\textbf{Limitation 2: Missing Alert Summary Process}}\label{limit2}
Even after denoising, individual systems can still generate numerous alerts that require further interpretation. In current practice, these alerts are listed one by one and missing an overarching narrative or aggregated view. 
Moreover, these individual alerts flag deviations from normal patterns without providing the essential context needed to understand their significance or root causes. 
This contextual vacuum forces SREs to engage in time-consuming investigations, manually correlating fragmented information from disparate sources, such as system documents, alert explanations, and incident tickets, to understand underlying issues. The heavy reliance on human expertise creates bottlenecks in the diagnosis process, delaying resolution. To address this, an automated alert summarization module is needed to translate cryptic alerts into actionable insights, leveraging internal knowledge (\eg system documents, alert rule explanations, incident tickets) to provide a coherent summarization explaining what the issue is, why it occurred, and how to resolve it.

\subsubsection{\textbf{Limitation 3: Missing Alert Rule Refinement Process}}\label{limit3}
In addition to the challenges encountered during alert handling, our current system suffers from limited alert rule management (§~\ref{problem2}). Once an alert rule is generated in \company, it is often considered valid indefinitely, lacking an iterative \emph{Alert Refinement} phase in its lifecycle. However, real-world systems frequently undergo changes in traffic patterns and operating environments, rendering static rules increasingly susceptible to persistent false alarms. 
Because SREs are often preoccupied with real-time production incidents, there is little opportunity to feed operational insights (\eg diagnosis result) back into rule design and maintenance, leading to a gradual decline in overall alert policy effectiveness.


\begin{figure}[t]
    \centering
    \includegraphics[width=1\linewidth]{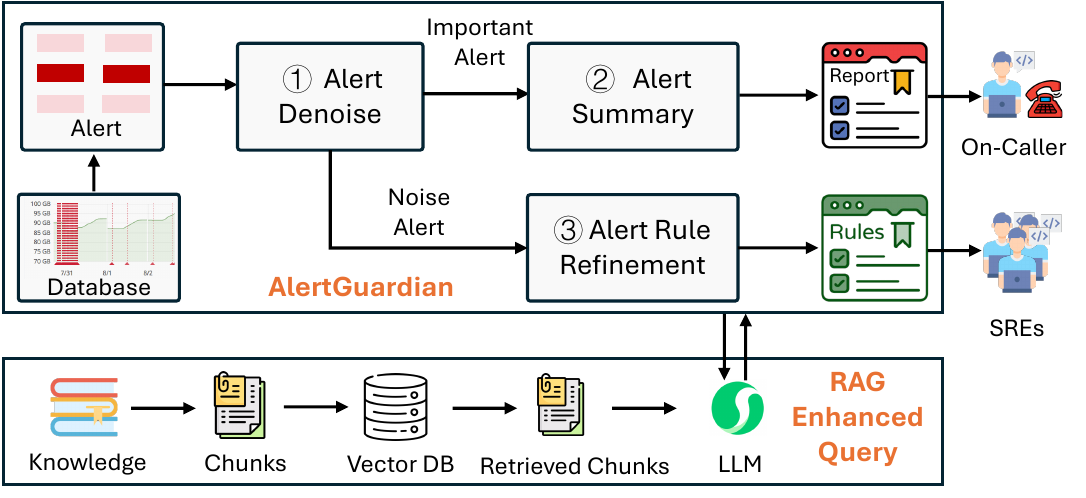}
    \vspace{-0.2in}
    \caption{Overview of \nm in \company.} 
    \vspace{-0.25in}
    \label{fig:framework}
\end{figure}

\section{Experience of Designing \nm}

To overcome the limitations of our alert system (\S\ref{sec:existing}), we propose a new alert system, named \nm, designed to manage the entire alert life-cycle. Fig.~\ref{fig:framework} shows the overview of \nm, comprising three key phases: 
\textcircled{\raisebox{-0.9pt}{1}} \textit{Alert Denoise} (\S~\ref{sec:denoise}) filters out high-frequency and periodic alerts based on a graph model, cutting down on noise in real-time. 
\textcircled{\raisebox{-0.9pt}{2}} For remaining important alerts, \textit{Alert Summary} (\S~\ref{sec:summary}) provides concise summaries (\eg potential causes and recommended solutions) for on-call engineers based on a RAG-enhanced LLM.
\textcircled{\raisebox{-0.9pt}{3}} For noisy alerts, \textit{Alert Rule Refinement} (\S~\ref{sec:refine}) optimizes rules for noisy alerts as an offline task based a RAG-enhanced LLM and multi-agent workflow, driving continuous improvement of alert rule quality.


\subsection{Alert Denoise}\label{sec:denoise}
To address the obsolete alert denoising mechanism identified in our existing system (\S\ref{limit1}), this study proposes a graph learning model to effectively handle the vast number of alert attributes and their increasingly diverse attribute value distributions. Instead of leveraging LLM, we opted for a traditional lightweight model for two key reasons: (1) Cost Efficiency: Our systems generate a massive number of alerts daily (\S\ref{problem1}), and using LLM would incur significant token costs, making it economically unfeasible. (2) Inference Speed: The inherent latency of LLMs, stemming from their autoregressive, token-by-token generation process and substantial computational overhead, presents a critical bottleneck~\cite{preserve}. When processing high-volume alert streams, this latency makes it infeasible to meet the stringent requirements of real-time diagnosis.

\subsubsection{Alert Preprocess}\label{sec:pre}
For the given alert dataset, we initially segment it into discrete 1-minute time windows to facilitate analysis. Notably, certain alert rules incorporate a pre-defined duration. For example, alerts triggered by Rule-1 in Fig.~\ref{fig:alertrule}, if unresolved, persist as active for 5 minutes post-firing. To accurately represent this persistence, we duplicated these alerts across the subsequent 5 time windows following their initial occurrence. To isolate noisy alerts from important ones, we introduce a \textit{virtual noisy alert}, fired every minute to emulate persistent noise patterns. The core intent behind mandating the virtual noisy alert's presence each minute is to pinpoint real alerts exhibiting high co-occurrence frequencies with it. These alerts are likely indicative of persistent noise (\eg false positives from heartbeat checks) that requires exclusion.

For time window $i$, we construct a relationship matrix $M_i$ to quantify the co-occurrence frequency among the alerts. If alert $u$ and alert $v$ co-occur in this window, the corresponding element $M_i[u,v]$ is marked as 1; otherwise, it is 0. 
Spanning the time windows from $1$ to $\tau$, we obtain a sequence of $\tau$ co-occurrence matrices, denoted $ \{M_1, M_2, \ldots, M_\tau\}$. These matrices are subsequently aggregated to form a statistical co-occurrence matrix $\mathcal{M}$, encapsulating the co-occurrence relationships among alerts over the entire period. This aggregation yields a sparse matrix $\mathcal{M}$, offering a concise representation of alert co-occurrence suitable for further analysis.

Subsequently, we seek to derive embeddings for the alerts to support advanced modeling efforts. Directly computing embeddings for all alert presents challenges due to their vast quantity. To address this, we compute embeddings for individual alert attributes and derive each alert's embedding by aggregating its constituent attribute embeddings. However, certain attributes, such as Pod ID, possess extensive value spaces and primarily function as entity identifiers rather than indicators of anomalous behavior. Directly embedding these attributes would necessitate processing an excessive number of values, thereby diluting the focus on features pertinent to anomalies. To counter this, we anonymize such identifier attributes by converting their values to a standardized format, ``ANON\_'' appended with the attribute name (\eg ``ANON\_POD\_ID''). This anonymization reduces the cardinality of attribute values, allowing the algorithm to prioritize computational resources on features directly associated with system anomalies.

\subsubsection{Graph Learning Model Training}
Alerts in a system exhibit complex temporal and content-based relationships~\cite{lin2018collaborative}.
\nm model temporal correlation using graph structures, where alerts are nodes, and virtual noisy nodes connect to all alerts to simulate persistent alerts.
For each node pair $(u, v)$, edge attributes are defined by co-occurrence frequency $k$ and total occurrence count $c$ across all time points.
To address content correlation, we adopt a unique attribute-value encoding approach to capture similarity among alerts. Unlike traditional bag-of-words encoding~\cite{almeida2019word}, we assign a unique code to each attribute-value pair without considering semantics, focusing on identical pairs. Specifically, we sequentially encode attribute-value pairs based on their dataset appearance order. For two alerts, if they share more identical attribute-value pair codes, they are considered more similar.

Our model GraphGuardian integrates Large-Scale Information Network Embedding (LINE)~\cite{tang2015line} with the Transformer architecture~\cite{transformer} to capture local and global alert relationships. LINE directly utilizes the co-occurrence frequency of alerts to learn embeddings. If two alert nodes are frequently directly connected (high co-occurrence count), their embedding vectors will be very close in space. This is local structural information based on ``first-order'' and ``second-order'' proximity. Transformer analyzes all attribute-value pair codes of an alert simultaneously through its self-attention mechanism. It is capable of learning more complex higher-order dependencies that go beyond simple pairwise co-occurrences.This ability allows it to understand the global contextual information within an alert.


For alert pairs $(u, v)$, embeddings $h_u$ and $h_v$ are generated, with similarity measured by the squared cosine distance:
\begin{equation}
d(h_u, h_v) = \left(1 - \frac{h_u \cdot h_v^T}{\|h_u\| \|h_v\|}\right)^2,
\label{eq:d}
\end{equation}
where distance $d(h_u, h_v) \to 0$ indicates high dissimilarity and $d(h_u, h_v) \to  1$ indicates high similarity.

To handle low-frequency alerts, we use maximum likelihood estimation (MLE) as the loss function, which is less sensitive to infrequent alerts than traditional methods (\eg mean squared error (MSE), mean absolute error (MAE)). MLE treats co-occurrence frequency as a binomial distribution, optimizing parameters by maximizing the likelihood:
\begin{equation}
\text{Binomial}(k, c, p) = \binom{c}{k} p^k (1-p)^{c-k},
\label{eq:binomial}
\end{equation}
where $p = d(h_u, h_v)$. We chose the binomial distribution because it is less sensitive to low-frequency events, which is very important in our sparse data scenario. The loss is the negative logarithm of the probability mass function:
\begin{equation}
\text{loss} = -\log (\text{Binomial}(k, c, p)).
\label{eq:loss}
\end{equation}

We use the Adam optimizer~\cite{kingma2014adam} for efficient parameter updates, combining stochastic gradient descent with adaptive learning rates and momentum. In each iteration, a random alert pair is processed, with the Transformer capturing global dependencies and LINE preserving local connectivity. Minimizing MLE loss improves model performance and generalization.

\begin{figure}[t]
    \centering
    \includegraphics[width=1\linewidth]{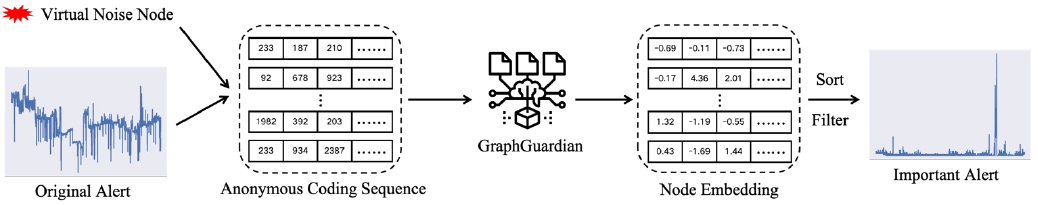}
    \vspace{-0.2in}
    \caption{Example of alert denoise process.} 
    \vspace{-0.25in}
    \label{fig:denoise_example}
\end{figure}

\subsubsection{Model inference}
During online inference, our graph-based model processes incoming alerts to distinguish critical alerts from noise in real-time. 
\nm measures the similarity between alert and virtual alerts based on node-to-node associations, as detailed in Eq.~\ref{eq:d}. The higher similarity indicates a higher likelihood of being a noise. 
To determine noisy alerts, we introduce a similarity threshold $\theta \in [0, 1]$. Alerts with $d(h_u, h_{v_{\text{noise}}}) \geq \theta$ are classified as noise, while others are deemed important. A higher $\theta$ increases precision by filtering more alerts, but risks missing critical ones, while a lower $\theta$ improves recall at the cost of retaining more noise. 
In practice, a default $\theta$ value of 0.7 is recommended. This value strikes a balance by classifying alerts with a moderately high similarity to the virtual noisy node as noise, while preserving alerts with a lower similarity as potentially critical. Fig.~\ref{fig:denoise_example} shows an example of an alert denoise process.

\deployment{A virtual persistent noisy alert identifies persistent noise patterns via co-occurrence analysis. Anonymizing high-cardinality attributes reduces computational overhead, focusing on anomaly features.}

\subsection{Alert Summary}\label{sec:summary}

Even after the denoising stage, \nm can still produce multiple critical alerts (\eg more than a dozen), posing a risk of overwhelming SREs.
As shown in Limitation 2 (\S~\ref{limit2}),  individual alerts flag deviations from normal patterns without providing the essential context needed to understand their implications or root causes. To address this limitation, \emph{Alert Summary} module employs RAG to incorporate internal knowledge (\eg system documents, alert rule explanations, incident tickets) of \company into an LLM, thereby generating concise yet actionable alert summaries (\eg fault explanations, localization, and resolutions).

\textbf{Why LLM?}  LLMs are sophisticated neural networks, extensively trained to comprehend and generate text with high contextual awareness. Their key advantage for alert contextualization lies in their natural language processing capabilities, which enable them to interpret technical alert data and translate it into coherent, human-readable narratives, highlighting both symptoms and potential causes. This bridge between machine-generated outputs and human operators significantly accelerates comprehension and response.

\textbf{Why Connect LLMs to RAG?}
LLMs have static, general-purpose knowledge, but lack access to private operational insights (\eg alert rule definitions or tailored troubleshooting steps) of \company. While fine-tuning an LLM for alert summary is possible, it demands significant computational resources, domain-specific datasets, and risks becoming outdated or overfitted. In contrast, RAG provides a straightforward alternative: it embeds relevant internal documents in a vector database and retrieves Top-K related excerpts to contextualize the LLM prompts dynamically~\cite{rag}. Because \company already hosts a mature RAG system for SRE queries, \nm simply reuses this existing infrastructure. Considering the real-time nature of diagnosis, we employ the general LLM model \emph{Deepseek V3}~\cite{DeepseekV3} instead of reasoning model for faster inference speed.

Once \textit{Alert Denoise} module (\S\ref{sec:denoise}) produces its refined alert set, \emph{Alert Summary} module proceeds as follows:
\begin{enumerate}[leftmargin=*, label=(\arabic*)]
    \item \textit{Per-Alert Retrieval}: For each critical alert in a one-minute window, a RAG query uses its attributes (\eg rule, component) to retrieve relevant chunks (\eg alert rule explanation, incident tickets). 
    \item \textit{Context Fusion}: Chunks are filtered for relevance and augmented with a cached system context summary (\eg recent system state).
    \item \textit{LLM Actionability and Summary}: LLM evaluates actionability across the batch, analyzing alerts, knowledge fragments, and relationships (\eg temporal or causal).  If an alert is deemed non-actionable (\eg transient anomaly), the system remains silent, requiring no SRE intervention. Conversely, actionable alerts yield a consolidated fault summary that describes the root cause, an explanation, a proposed solution, and references to pertinent knowledge.
\end{enumerate}
This pipeline applies chain-of-thought reasoning~\cite{COT}. Upon determining an alert to be actionable, the module outputs a structured report encompassing:
\begin{itemize}[leftmargin=*]
\item \textbf{Alerts.} Lists up to five alerts most relevant to the current diagnostic process.
\item \textbf{Root Cause.}  Root cause component leading to the anomaly.
\item \textbf{Explanation.} A rationale underpinning the fault diagnosis.
\item \textbf{Solution.} Concrete remediation steps for the root cause.
\item \textbf{Reference.} Pertinent knowledge link to documents.
\end{itemize}

\deployment{By leveraging RAG,  Alert Summary dynamically retrieves internal knowledge, enabling context-aware summaries without costly fine-tuning.}


\begin{figure}[t]
	\centering
	\includegraphics[width=0.5\textwidth]{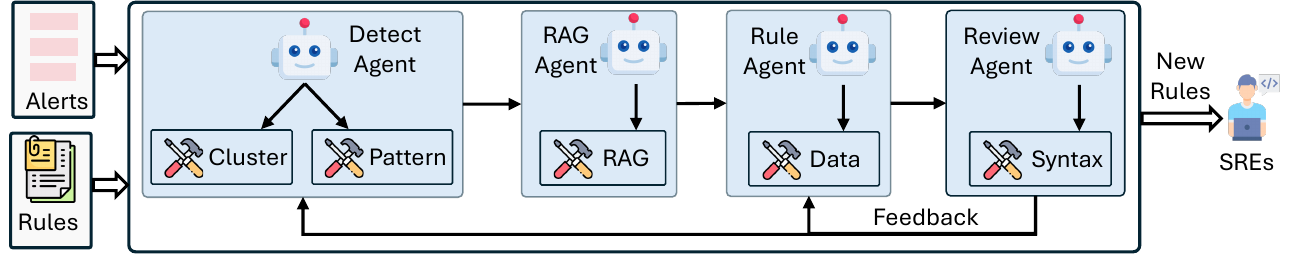}
	\vspace{-0.2in}
	\caption{Multi-agent workflow of Alert Rule Refinement.}
 	\label{fig:refine}
 	\vspace{-0.25in}
\end{figure}

\subsection{Alert Rule Refinement}\label{sec:refine}

To overcome Limitation 3 (\S\ref{limit3}),  \textit{Alert Rule Refinement} module optimizes rules responsible for noise alerts identified by denoising module (\S\ref{sec:denoise}). This process reduces recurring noise while maintaining sensitivity to critical alerts. Rather than creating new rules, which could introduce syntax errors or inaccuracies due to LLM randomness~\cite{gu2025argosagentictimeseriesanomaly}, the module refines existing rules using the LLM capacity, leveraging alert context and current rule definitions for enhanced precision.

\textbf{Multi-Agent Workflow.} As shown in Fig.~\ref{fig:refine}, \nm introduces a multi-agent workflow to refine alert rules, executed every large time window (30 minutes by default) to balance efficiency and effectiveness. Agents collaborate in a pipeline without a central orchestrator, ensuring streamlined communication. Within a time window, all alerts triggered by a rule are aggregated and based on the denoising module’s results, they are classified into noise and critical alerts. For each rule, the workflow is structured as follows:
\begin{enumerate}[leftmargin=*, label=(\arabic*)]
\item Detect Agent is used to identify noise patterns with two specialized tools: (i) Clustering Tool applies the HDBSCAN algorithm~\cite{campello2015hierarchical} to group noise alerts by attributes (\eg rule name, explanation, attributes), detecting patterns like semantically similar alerts without predefined cluster counts.
(ii) Periodicity Pattern Tool identifies persistent or periodic alerts (\eg daily CPU spikes) based on frequency and timing. Detect Agent categorizes alerts into clusters or patterns requiring refinement.
\item RAG Agent queries the RAG Vector DB for each categorized group, aggregating key attributes to retrieve relevant knowledge chunks (\eg rule explanation). 
\item Rule Agent: Refines rules by analyzing categorized alerts and retrieved chunks, employing four policies:
\begin{itemize}
    \item \textit{Rule Deduplication}: Identifies and merges redundant rules by comparing their semantic intent and operational overlap, using LLM-driven natural language understanding to ensure functional equivalence while simplifying the rule set.
    \item \textit{Rule Aggregation}: Consolidates rules targeting similar alerts (\eg Rule-1 and Rule-2 in Fig.~\ref{fig:alertrule} both focus on pod not-ready status) by generalizing attributes and introducing conjunctive conditions (\eg AND logic). The LLM infers common patterns across alerts and proposes unified rules that reduce redundancy while maintaining coverage.
    \item \textit{Threshold Adjustment}: Analyzes statistical distributions of metrics (\eg CPU usage, latency) from the knowledge base, using LLM to recommend adaptive thresholds that align with traffic patterns, seasonal variations, or anomaly trends. For instance, it would adjust a CPU alert threshold from 40\% to 60\% in Fig.~\ref{fig:alertrule} based on historical data correlations, suppressing noise without missing critical events.
    \item \textit{Temporal Analysis}: Incorporates time-based conditions for recurring patterns (\eg excluding maintenance windows or silencing daily spikes). The LLM enhances this by predicting optimal time windows from contextual cues and suggests precise temporal filters (\eg ``trigger only if sustained for 5 minutes'').
\end{itemize}
The Rule Agent’s LLM also generates detailed rationales for each refinement (\eg ``Threshold raised to 60\% based on 95th percentile of last 30 days’ data''), enhancing transparency and trust.
\item Review Agent validates refined rules using a syntax checking tool to ensure rule integrity and a simulation tool that tests rules against historical alerts, confirming noise reduction without compromising critical alerts. If validation fails, the Review Agent triggers iterative refinement.
\end{enumerate}

\textbf{Iterative Feedback Loop.} The Detect Agent, Rule Agent, and Review Agent form a feedback loop to iteratively refine alert rules until they meet precise stopping conditions, ensuring reliability and effectiveness. The loop terminates when all the following criteria are satisfied:
\begin{itemize}[leftmargin=*]
    \item \textbf{Syntax Integrity}: Refined rules must be free of syntax errors, verified by the Review Agent’s syntax checking tool, ensuring they are executable by the alert system.
    \item \textbf{Preservation of Critical Alerts}: Refined rules must retain all critical alerts, confirmed by the Review Agent’s simulation tool, which tests rules against historical data to ensure no critical alerts are suppressed.
    \item \textbf{Noise Reduction Threshold}: The noise alert ratio (proportion of non-critical alerts) must fall below a predefined threshold (5\% by default), as evaluated by the Review Agent’s simulation tool against historical data, while maintaining full coverage of critical alerts.
\end{itemize}
During each iteration, the Review Agent assesses the refined rules against these criteria using inputs from the Detect Agent. If any condition is not met, such as syntax errors, loss of critical alerts, or a noise ratio above the threshold, the Review Agent provides targeted feedback to the Rule Agent, prompting further refinement. This process continues until all stopping conditions are satisfied, ensuring robust optimization. 
If the above conditions are not met after 30 iterations, the optimization process for the rule is terminated and the rule remains unoptimized to prevent excessive computational overhead.
The refined rules are then submitted to the SREs for final approval, maintaining a human-in-the-loop framework to ensure operational trust and accountability. 

\deployment{The iterative feedback loop and LLM-generated rationales provide SREs with transparent insights, boosting confidence in rule adjustments.  }


\section{Experimental Evaluation} \label{sec:exp}
To evaluate the effectiveness of \nm, we conducted an experimental study aimed at addressing the following research questions (RQ):
\begin{itemize}[leftmargin=*]
    \item \textbf{RQ1:} How does \nm  perform in alert denoise?
    \item \textbf{RQ2:} How does \nm perform in alert summary?
    \item \textbf{RQ3:} How well does \nm refine alert rules?
\end{itemize}

\subsection{Experiment Setup}

\textbf{Dataset.} We collected alert and alert rule data from four large-scale cloud systems at \company, supporting gaming, office, media, and education services. As shwon in Table~\ref{tab:dataset}, these datasets, denoted $\mathcal{A}, \mathcal{B}, \mathcal{C}$, and $\mathcal{D}$, each span 9 days and encompass thousands of services, generating hundreds of thousands of alerts daily. Alerts and alert rules in \company conform to the Prometheus generic alarm system format, ensuring the generalizability of our approach. Each dataset includes over 100 incidents, with annotations for critical alerts and root cause components. Table~\ref{tab:dataset} summarizes the dataset characteristics.

\begin{table}
    \centering
    \caption{Detailed information on production datasets.}
    \vspace{-0.05in}
    \label{tab:dataset}
    \resizebox{0.9\linewidth}{!}{\begin{tabular}{c|c|c|c|c}
    \toprule 
    \textbf{Dataset} & \textbf{Domain} & \textbf{\#Rule} & \textbf{\#Alert} & \textbf{\#Incident} \\ 
    \midrule 
    $\mathcal{A}$ & Game & 12,960 & 2,853,345 & 138 \\
    $\mathcal{B}$ & Office & 3,544 & 1,243,259 & 114 \\
    $\mathcal{C}$ & Media & 59,607 & 3,883,293 & 187 \\
    $\mathcal{D}$ & Education & 6,962 & 2,692,964 & 179 \\
    \bottomrule  
    \end{tabular}}
    \vspace{-0.2in}
\end{table}


\textbf{Implementation.} We implemented \nm using Python 3.7.2 and PyTorch 1.13.1~\cite{paszke2019pytorch}. The graph learning model for alert denoising was trained on alerts from the first 6 days of each dataset and evaluated on the final 3 days to ensure temporal separation between training and testing. For alert summarization and rule refinement, we leveraged \company's existing RAG infrastructure, which stores operational knowledge in a vector database. The RAG system, widely adopted by 86\% of enterprises deploying LLMs~\cite{K2view}, enhances generalizability. 
We use Deepseek V3~\cite{DeepseekV3} in alert summary and use Deepseek R1~\cite{DeepseekR1} in alert rule refinement provided by \company by default.
All experiments were conducted on a server running TencentOS Server 3.2, equipped with a 24-core Intel Xeon E5-2690 v3 CPU, 128 GB of RAM, and an NVIDIA Tesla V100 GPU.






\subsection{RQ1: Performance in Alert Denoise}

We evaluated the effectiveness of \nm in alert denoising from the following three aspects:
\begin{itemize}[leftmargin=*]
    \item \textbf{RQ1.1:} What is the extent of alert volume reduction achieved by \nm?
    \item \textbf{RQ1.2:} How accurately does \nm retain critical alerts for identifying failures?
    \item \textbf{RQ1.3:} How efficient is the alert denoise process?
\end{itemize}

\textbf{Evaluation metrics.}
For RQ1.1, we quantified alert reduction using the reduction ratio, defined as:  $\frac{N - M}{N}$, where $N$ is the initial number of alerts, and $M$ is the number of alerts remaining after denoising. This metric reflects the reduction in alert burden for SREs. For RQ1.2, we assessed the accuracy of retaining critical alerts using Precision (P), Recall (R), and F1-score (F1). Precision is the proportion of critical alerts among the retained alerts, Recall is the proportion of true critical alerts retained relative to all true critical alerts, and F1-score is their harmonic mean, providing a balanced measure of accuracy. Critical alerts were derived from SRE-provided incident reports, serving as ground truth.
For RQ1.3, we measured computational efficiency by recording the average inference time per alert batch (processed in 1-minute windows) on the experimental hardware.

\textbf{Baselines.} We compared \nm against the following baseline methods to contextualize its performance:
\begin{itemize}[leftmargin=*]
    \item Severity-based method is a common heuristic where only alerts with high severity (\eg critical level in our datasets) are prioritized, reflecting typical SRE practices.
    \item Cluster-based method UHAS~\cite{zhao2020understanding} uses Extreme Value Theory (EVT)~\cite{siffer2017anomaly} and Isolation Forest~\cite{liu2008isolation} for clustering, retaining only alerts corresponding to cluster centroids.
    \item Window-based method OAS~\cite{chen2022online}  suppresses a new alert with $A_{i}^{t_1}$ if it is sufficiently similar to a precedent alert $A_{i}^{t_0}$ and $t_1-t_0 < \omega $. We set the window size $\omega$ to 1 minutes.
    \item \nm without Anonymization (\nm w/o Non): A variant of \nm without attribute anonymization, to evaluate the impact of anonymization on denoising performance.
\end{itemize}

\begin{table}
\centering
\caption{Comparison of alert reduction ratios.}
\vspace{-0.05in}
\label{tab:reduce}
\resizebox{0.9\linewidth}{!}{
    \begin{tabular}{@{}lcccc@{}}
     \toprule
     \multirow{2}{*}{\begin{tabular}[c]{@{}c@{}}\textbf{Method}\end{tabular}} & \multicolumn{4}{c}{\textbf{Alert Reduction Ratio(\%)}} \\ \cmidrule(l){2-5} 
                                                                            & $\mathcal{A}$ & $\mathcal{B}$ & $\mathcal{C}$ & $\mathcal{D}$   \\ \midrule
    \nm & \textbf{95.10}  & \textbf{93.82}  & \textbf{95.50}  & \textbf{95.00}  \\
    \nm w/o Anon & 90.20  & 88.00  & 90.00  & 89.00  \\
    Severity & 85.04  & 83.92  & 85.00  & 84.00  \\
    OAS~\cite{chen2022online} & 88.00  & 87.00  & 88.00  & 87.00  \\
    UHAS~\cite{zhao2020understanding} & 91.00  & 89.00 & 91.00  & 90.00  \\
    \bottomrule  
    \end{tabular}}
\vspace{-0.2in}
\end{table}

\textbf{RQ1.1: Effort Reduction.} 
The experimental results in Table.~\ref{tab:reduce} show that \nm achieves the highest alert reduction ratios across four cloud systems, ranging from 93.82\% to 95.50\%, significantly outperforming all baseline methods and demonstrating its superior ability to filter noise. Its success is attributed to the graph learning model and attribute anonymization techniques. In contrast, \nm w/o Anon, without anonymization, exhibits a 5–6\% reduction ratio drop (\eg from 95.10\% to 90.20\% in dataset $\mathcal{A}$), highlighting anonymization’s importance for handling high-cardinality attributes like Pod IDs. Among baselines, severity-based method performs worst due to its reliance on simplistic heuristics, while UHAS and OAS show moderate performance but lack precision. 

\begin{figure*}[t]
	\centering
	\includegraphics[width=1\textwidth]{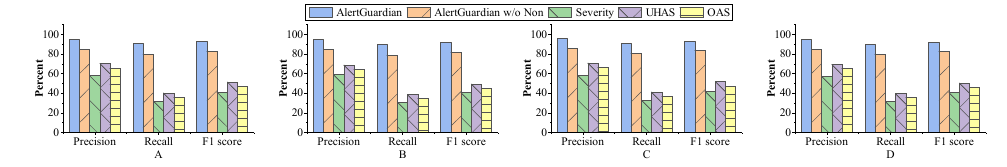}
    \vspace{-0.3in}
    \caption{Comparison of alert denoise accuracy.} 
    \vspace{-0.2in}
    \label{fig:precision}
\end{figure*}

\textbf{RQ1.2: Denoise Accuracy.}
Figure~\ref{fig:precision} shows the denoise accuracy of \nm, \nm w/o Anon, UHAS~\cite{zhao2020understanding}, OAS~\cite{chen2022online}, and Severity-Based across four datasets using Precision, Recall, and F1 Score.
As shown in Fig.~\ref{fig:precision}, UHAS and OAS achieve high precision but low recall, as their clustering (UHAS) and window-based suppression (OAS) mechanisms discard some critical alerts, limiting their effectiveness for fault diagnosis. In contrast, \nm outperforms both, yielding higher precision and recall while retaining fewer alerts, enhancing fault diagnosis efficiency in large-scale systems. The \nm w/o Anon variant shows reduced performance, underscoring the critical role of anonymization in handling high-cardinality attributes and improving noise filtering. These results highlight the superior balance of precision and recall, making it a robust solution for reliable alert management.

\textbf{RQ1.3: Efficiency of Alert Denoise.}
Our objective is to swiftly handle online alert storms and assist SREs in fault remediation, necessitating a highly efficient, low-latency system.
During the offline training phase, we employ a distributed framework (\ie Ray~\cite{moritz2018ray}) with data parallelism. The training complexity is approximately $O(N^2 \times M)$, where $N$ is the number of alert entities and $M$ is the number of training epochs. This process is highly efficient in practice: training on a dataset with one million alert ($N=10^6$) takes only 20 minutes, and for a typical cloud system with data from the last seven days (under five million alerts), training completes within 100 minutes.
For online inference, the alert volume for a single system per minute usually stays below 1000. The computational complexity for this stage is approximately $O(N^2)$, enabling the system to perform alert denoising and summary extraction in under 200 milliseconds. Furthermore, response times can be further optimized through techniques like caching for accelerated data access.

\subsection{RQ2: Performance in Alert Summary}

To evaluate the effectiveness of alert summary results, we conducted experiments without labeled summary data, combining quantitative and qualitative metrics. Quantitative metrics included Action/No-Action Accuracy (accuracy in classifying summaries as requiring action or not, validated against incident reports) and Root Cause Analysis Accuracy (accuracy in identifying root causes, assessed via heuristic matching with incident report annotations). Qualitative assessment involved two SREs scoring summaries for Actionability and Relevance (1–5 scale). Due to the high cost of manual labeling, we focused solely on Dataset $\mathcal{A}$. We compared \nm (RAG + Deepseek V3) against two 
methods OAS~\cite{chen2022online} and UHAS~\cite{zhao2020understanding} that are not based on LLM and two methods \nm w/o RAG and \nm (RAG + Qwen 2.5 72B) that are based on LLM.

\begin{table}[t]
\centering
\caption{Comparison of alert summary performance.}
\vspace{-0.05in}
\label{tab:summary}
\resizebox{0.9\columnwidth}{!}{
\begin{tabular}{lcccc}
\toprule
\textbf{Method} & \textbf{Action Acc.(\%)} & \textbf{RCA(\%)} & \textbf{Action} & \textbf{Rele.} \\
\midrule
\nm (DeepSeek) & \textbf{98.5} & \textbf{90.5} & \textbf{4.8} & \textbf{4.9} \\
\nm w/o RAG & 86.5 & 82.5 & 3.0 & 3.1 \\
\nm (Qwen) & 91.5 & 88.0 & 4.0 & 4.5 \\
OAS~\cite{chen2022online} & 70.0 & 64.5 & - & - \\
UHAS~\cite{zhao2020understanding}  & 72.5 & 67.0 & - & - \\
\bottomrule
\end{tabular}}
\vspace{-0.25in}
\end{table}

As shown in Table~\ref{tab:summary}, \nm achieves the best performance, with Action Accuracy of 98.5\%, RCA Accuracy of 90.5\%, Actionability of 4.8, and Relevance of 4.9. Its RAG-based approach with Deepseek V3 leverages retrieved context to produce concise, actionable summaries that effectively identify critical alerts and root causes. The \nm with Qwen 2.5 72B baseline performs well but is limited by the smaller model size of Qwen 2.5 72B, reducing its capacity for nuanced summarization. \nm w/o RAG shows significantly lower performance, as the lack of background knowledge hampers Deepseek V3’s ability to contextualize alerts. Non-LLM baselines, UHAS and OAS, yield poor quantitative results and are incapable of generating interpretable summaries, rendering Actionability and Relevance scores inapplicable. Their clustering and window-based mechanisms oversimplify alert relationships, missing critical fault details. The superior quantitative and qualitative performance demonstrates its efficacy in reducing alert fatigue while enhancing fault diagnosis in large-scale cloud systems.

\subsection{RQ3: Performance in Alert Rule Refinement}

As described in \S\ref{sec:refine}, \nm employs four policies: Rule Deduplication, Rule Aggregation, Threshold Adjustment, and Temporal Analysis. New rules are recommended to SREs, who assess their acceptance. Table~\ref{tab:refilne} presents the performance of \nm, recommending 221–375 rules across datasets, with 73–117 accepted (accept rates 31.2–33.0\%). Rule Deduplication achieves the highest accept rates (80.0–83.3\%) due to its clear identification of redundant rules. Using LLM-driven semantic analysis, \nm detects overlapping rules, offering low-risk simplifications that SREs readily accept to reduce noise. Rule Aggregation and Threshold Adjustment yields moderate accept rates, as consolidating similar alerts and adjusting threshold risks overgeneralization, requiring SRE validation. Temporal Analysis records the lowest accept rates, as time-based conditions are error-prone in variable workloads.

\begin{table}[t]
\centering
\caption{Rule refinement performance of \nm.}
\vspace{-0.05in}
\label{tab:refilne}
\resizebox{0.9\columnwidth}{!}{
\begin{tabular}{ccccc}
\toprule
\textbf{Dataset} & \textbf{Policy} & \textbf{Recommend} & \textbf{Accept} & \textbf{Accept Rate(\%)} \\
\midrule
\multirow{5}{*}{\(\mathcal{A}\)} & Rule Deduplication & 50 & 40 & 80.0 \\
& Rule Aggregation & 100 & 28 & 28.0 \\
& Threshold Adjustment & 80 & 20 & 25.0 \\
& Temporal Analysis & 75 & 8 & 10.7 \\
\cmidrule{2-5}
& \textbf{Total} & \textbf{305} & \textbf{96} & \textbf{31.5} \\
\midrule
\multirow{5}{*}{\(\mathcal{B}\)} & Rule Deduplication & 36 & 30 & 83.3 \\
& Rule Aggregation & 75 & 22 & 29.3 \\
& Threshold Adjustment & 60 & 14 & 23.3 \\
& Temporal Analysis & 50 & 7 & 14.0 \\
\cmidrule{2-5}
& \textbf{Total} & \textbf{221} & \textbf{73} & \textbf{33.0} \\
\midrule
\multirow{5}{*}{\(\mathcal{C}\)} & Rule Deduplication & 62 & 50 & 80.6 \\
& Rule Aggregation & 120 & 35 & 29.2 \\
& Threshold Adjustment & 100 & 25 & 25.0 \\
& Temporal Analysis & 93 & 7 & 7.5 \\
\cmidrule{2-5}
& \textbf{Total} & \textbf{375} & \textbf{117} & \textbf{31.2} \\
\midrule
\multirow{5}{*}{\(\mathcal{D}\)} & Rule Deduplication & 46 & 38 & 82.6 \\
& Rule Aggregation & 90 & 26 & 28.9 \\
& Threshold Adjustment & 70 & 17 & 24.3 \\
& Temporal Analysis & 67 & 8 & 11.9 \\
\cmidrule{2-5}
& \textbf{Total} & \textbf{273} & \textbf{89} & \textbf{32.6} \\
\bottomrule
\end{tabular}}
\vspace{-0.2in}
\end{table}

\section{Discussion}

\subsection{Success Stories}

In February~2024, alert denoise module of \nm was deployed at \company, remaining in stable operation for over a year. Its alert summary and rule refinement modules have been in pilot use for more than three months in System~\textit{A}, one of \company’s major gaming platforms supporting tens of millions of users in real-time multiplayer game sessions. Prior to implementing \nm, System~\textit{A} faced severe alert storms: over 30,0000 alerts were generated daily from more than ten thousand rules.

\nm effectively addresses the alert storms with three core features, significantly improving both system reliability and operational efficiency:
\begin{itemize}[leftmargin=*]
    \item \textbf{Alert Denoise.}  Leveraging graph-based learning and attribute anonymization, \nm reduces the system’s daily alerts by 95\%, from 300,000 down to about 15,000 per day (averaging 10 alerts per minute). Meanwhile, critical alerts maintain an F1-score of 0.92, enabling downstream components to focus on pivotal alerts and filter out a large volume of irrelevant noise.
    \item \textbf{Alert Summary.}  For the alerts remaining in each one-minute window after denoising, alert summary module determines whether SRE intervention is required. If not, the system remains silent; otherwise, it clusters key alerts, identifies root causes, generates succinct summaries, and recommends solutions. This action/non-action decision achieves 98\% accuracy, substantially alleviating  ``alert fatigue'' among SREs. The mean time to recovery (MTTR) dropped from an average of 156 minutes to 21 minutes, improving efficiency by a factor of 7.4.
    \item \textbf{Alert Rule Refinement.}  By analyzing noisy alerts flagged by the alert denoise module, \nm generated over 300 rule optimization proposals, with near 100 adopted by SREs. Through continuous refinement, the system eliminates over 50,000 false positives per day.
\end{itemize}

\subsection{Lessons Learned}
The development and deployment of \nm at \company have provided valuable insights into effective alert life-cycle management for large-scale cloud systems. 

\textbf{Hybrid Model Approach Balances Efficiency and Contextual Depth.} \nm combines a lightweight graph learning model for alert denoising with a LLM for alert summarization, optimizing both computational efficiency and narrative richness. The small model leverages graph-based techniques to process high-volume alert streams, filtering noise while preserving critical signals. In contrast, the LLM enhances summarization by generating human-readable narratives that contextualize alerts with system dependencies and operational insights. This hybrid strategy demonstrates that integrating small models for scalable processing with LLMs for advanced semantic understanding can address diverse requirements in alert management effectively.

\textbf{RAG-Enabled Knowledge Integration Enhances Scalability.} By employing RAG, \nm seamlessly incorporates private system documentation and operational knowledge without the need for resource-intensive fine-tuning. RAG retrieves relevant context, such as alert rule explanations and incident records, enabling the LLM to produce informed summaries that align with the organization’s proprietary environment. This approach highlights the power of RAG as a cost-effective method to enhance alert contextualization, making it adaptable to domain-specific needs in complex cloud systems.

\textbf{Iterative Feedback Loop Drives Rule Optimization.} The rule refinement process relies on an iterative feedback loop that continuously improves alert rules by incorporating insights from noise detection and validation. Agents analyze noise patterns and propose rule optimizations, such as deduplication or threshold adjustments, with LLM-generated rationales for transparency. If proposed refinements fail to reduce noise or suppress critical alerts, feedback triggers further iterations until accuracy is achieved. Final proposals are validated by SREs, ensuring a human-in-the-loop approach. This mechanism underscores the importance of iterative refinement to adapt rules to evolving system dynamics, maintaining long-term effectiveness with minimal manual intervention.

\subsection{Threats to Validity}
\textbf{Internal Threats.} Potential biases in data labeling could affect evaluation metrics. To mitigate this, we relied on rigorous incident reports and annotations validated by multiple SREs, ensuring high-quality ground truth. Additionally, LLM hallucinations posed a risk to summarization and rule refinement. We addressed this by integrating RAG to anchor outputs in verified knowledge and employing iterative feedback loops to refine LLM-generated content, enhancing reliability.

\textbf{External Threats.} The datasets, sourced exclusively from \company, may limit generalizability. However, we mitigated this by including four diverse domains (\ie gaming, office, media, and education) demonstrating the effectiveness of \nm across varied workloads. Furthermore, \company’s use of Prometheus-based alert rules~\cite{prometheus}, a de facto industry standard widely adopted in cloud monitoring, ensures strong compatibility and applicability to other systems, bolstering the framework’s external validity.

\vspace{-0.1in}
\section{Related Work}

\textbf{Alert Rule Management.} Major cloud providers, such as Azure~\cite{azurerule}, Google Cloud~\cite{googlerule}, and open-source tools like Prometheus~\cite{prometheus} and Grafana~\cite{grafana}, offer robust alert rule management capabilities, enabling users to create, modify, and delete rules. These systems provide flexible interfaces for defining thresholds and conditions but lack intelligent optimization mechanisms to adapt rules dynamically.  DEAR~\cite{DEAR} proposes evaluating existing rules to balance high accuracy and low network traffic volume without administrative overhead. However, DEAR primarily focuses on optimizing traffic transmission rather than enhancing rule quality for alert reduction. In contrast, \nm is the first to leverage LLMs for intelligent rule refinement, enabling iterative rule improvement.

\textbf{Alert Management.} Alert storms in large-scale cloud systems have been widely studied, with several approaches proposed to mitigate their impact on SREs~\cite{man2012alert,xu2017lightweight,chen2022online,mahdavi2020real,zhao2020understanding,zhao2020automatically}. Alert ranking methods~\cite{jiang2009ranking,tang2012optimizing,zhao2020automatically} rank alerts by severity, directing SREs to focus on the most critical issues first. Other works employ alert correlation and aggregation techniques~\cite{man2012alert,xu2017lightweight,chen2022online,lin2014unveiling}, grouping related alerts into single incidents to reduce the number of actionable items. Additionally, alert denoising approaches~\cite{aminanto2019automated,zhao2020understanding} identify and suppress noise alerts to alleviate SRE burden. 
However, these methods assume fixed alert attributes, failing to handle real-world scenarios where attribute counts vary due to system complexity.
COLA~\cite{COLA2024ICSE} leverages LLMs’ natural language understanding to group related alerts. However, COLA does not address the attribute combination explosion and requires substantial labeled data for supervised fine-tuning (SFT), which is often scarce in real-world alerts.
\nm addresses these gaps with an unsupervised framework that accommodates variable attributes. Furthermore, by integrating RAG-enhanced LLM summarization, our approach provides actionable narratives for critical alerts, accelerating RCA.

\vspace{-0.1in}
\section{Conclusion}
In this paper, we tackle the challenge of alert storms from the alert life-cycle management perspective. We introduce \nm, a novel framework that combines lightweight graph models with LLMs to optimize the entire alert life-cycle. Its central idea is twofold: On the online side, \nm first identifies and filters out noisy alerts, then leverages internal system knowledge to analyze and summarize the remaining critical alerts based on LLMs for rapid fault diagnosis. On the offline side, \nm applies LLM-based rule refinement to address the underlying causes of noisy alerts, ultimately preventing recurring alert storms at their source. This integrated strategy not only reduces the immediate burden of alert overload but also improves long-term system maintainability.

\section{Acknowledge}
We greatly appreciate the insightful feedback from the anonymous reviewers. We thank all SREs in the experiments for their feedback. This work was supported in part by National Key Research and Development Program of China (Grant Number: 2024YFB4505904), the National Natural Science Foundation of China under Grant 62272495 and the Guangdong Basic and Applied Basic Research Foundation under Grant 2023B1515020054. This work was also sponsored by Tencent. The corresponding author is Pengfei Chen.

\balance
\normalem
\bibliographystyle{IEEEtran}
\bibliography{references}

@inproceedings{chen2022online,
  title={Online summarizing alerts through semantic and behavior information},
  author={Chen, Jia and Wang, Peng and Wang, Wei},
  booktitle={ICSE 2022},
  pages={1646--1657},
  url          = {https://doi.org/10.1145/3510003.3510055},
  doi          = {10.1145/3510003.3510055},
  year={2022}
}

@inproceedings{zhao2020automatically,
  title={Automatically and adaptively identifying severe alerts for online service systems},
  author={Zhao, Nengwen and Jin, Panshi and Wang, Lixin and Yang, Xiaoqin and Liu, Rong and Zhang, Wenchi and Sui, Kaixin and Pei, Dan},
  booktitle={INFOCOM 2020},
  pages={2420--2429},
  year={2020},
  url          = {https://doi.org/10.1109/INFOCOM41043.2020.9155219},
  doi          = {10.1109/INFOCOM41043.2020.9155219},
  organization={IEEE}

}

@inproceedings{lin2014unveiling,
  title={Unveiling clusters of events for alert and incident management in large-scale enterprise it},
  author={Lin, Derek and Raghu, Rashmi and Ramamurthy, Vivek and Yu, Jin and Radhakrishnan, Regunathan and Fernandez, Joseph},
  booktitle={Proceedings of the 20th ACM SIGKDD international conference on Knowledge discovery and data mining},
  pages={1630--1639},
  url          = {https://doi.org/10.1145/2623330.2623360},
  doi          = {10.1145/2623330.2623360},
  year={2014}
}

@inproceedings{chendynamic,
  title={Dynamic Graph Neural Networks-based Alert Link Prediction for Online Service Systems},
  author={Chen, Yiru and Zhang, Chenxi and Dong, Zhen and Yang, Dingyu and Peng, Xin and Ou, Jiayu and Yang, Hong and Wu, Zheshun and Qu, Xiaojun and Li, Wei},
  booktitle    = {{ASE} 2023},
  pages        = {79--90},
  publisher    = {{IEEE}},
  year         = {2023},
  url          = {https://doi.org/10.1109/ASE56229.2023.00177},
  doi          = {10.1109/ASE56229.2023.00177},
}

@inproceedings{zhao2020understanding,
  title={Understanding and handling alert storm for online service systems},
  author={Zhao, Nengwen and Chen, Junjie and Peng, Xiao and Wang, Honglin and Wu, Xinya and Zhang, Yuanzong and Chen, Zikai and Zheng, Xiangzhong and Nie, Xiaohui and Wang, Gang and others},
  booktitle={ICSE-SEIP 2020},
  pages        = {162--171},
  publisher    = {{ACM}},
  year         = {2020},
  url          = {https://doi.org/10.1145/3377813.3381363},
  doi          = {10.1145/3377813.3381363},
}

@inproceedings{DEAR,
  author       = {Mathias Mormul and
                  Pascal Hirmer and
                  Christoph Stach and
                  Bernhard Mitschang},
  title        = {{DEAR:} Distributed Evaluation of Alerting Rules},
  booktitle    = {{CLOUD} 2020},
  pages        = {158--165},
  publisher    = {{IEEE}},
  year         = {2020},
  url          = {https://doi.org/10.1109/CLOUD49709.2020.00034},
  doi          = {10.1109/CLOUD49709.2020.00034},
}

@article{turgeman2022context,
  title={Context-aware incremental clustering of alerts in monitoring systems},
  author={Turgeman, Lior and Avrashi, Yaniv and Vagner, Gabriella and Azaizah, Nadeem and Katkar, Someshwar},
  journal={Expert Systems with Applications},
  volume={210},
  pages={118489},
  year={2022},
  url          = {https://doi.org/10.1016/j.eswa.2022.118489},
  doi          = {10.1016/J.ESWA.2022.118489},
  publisher={Elsevier}
}

@inproceedings{chen2021graph,
  title={Graph-based incident aggregation for large-scale online service systems},
  author={Chen, Zhuangbin and Liu, Jinyang and Su, Yuxin and Zhang, Hongyu and Wen, Xuemin and Ling, Xiao and Yang, Yongqiang and Lyu, Michael R},
  booktitle={ASE 2021},
  pages={430--442},
  year={2021},
  url          = {https://doi.org/10.1109/ASE51524.2021.9678746},
  doi          = {10.1109/ASE51524.2021.9678746},
  organization={IEEE}
}

@inproceedings{xu2017lightweight,
  title={Lightweight and adaptive service api performance monitoring in highly dynamic cloud environment},
  author={Xu, Jingmin and Wang, Yuan and Chen, Pengfei and Wang, Ping},
  booktitle={SCC 2017},
  pages={35--43},
  year={2017},
  url          = {https://doi.org/10.1109/SCC.2017.80},
  doi          = {10.1109/SCC.2017.80},
  organization={IEEE}
}

@article{man2012alert,
  title={An alert aggregation algorithm based on iterative self-organization},
  author={Man, Dapeng and Yang, Wu and Wang, Wei and Xuan, Shichang},
  journal={Procedia Engineering},
  volume={29},
  pages={3033--3038},
  doi = {https://doi.org/10.1016/j.proeng.2012.01.435},
  url = {https://www.sciencedirect.com/science/article/pii/S1877705812004456},
  year={2012},
  publisher={Elsevier}
}

@article{mahdavi2020real,
  title={A real-time alert correlation method based on code-books for intrusion detection systems},
  author={Mahdavi, Ehsan and Fanian, Ali and Amini, Fatima},
  journal={Computers \& Security},
  volume={89},
  pages={101661},
  year={2020},
  url          = {https://doi.org/10.1016/j.cose.2019.101661},
  publisher={Elsevier}
}

@inproceedings{jiang2009ranking,
  title={Ranking the importance of alerts for problem determination in large computer systems},
  author={Jiang, Guofei and Chen, Haifeng and Yoshihira, Kenji and Saxena, Akhilesh},
  booktitle={ICAC 2009},
  pages={3--12},
  url          = {https://doi.org/10.1016/j.cose.2014.12.003},
  year={2009}
}

@inproceedings{tang2012optimizing,
  title={Optimizing system monitoring configurations for non-actionable alerts},
  author={Tang, Liang and Li, Tao and Pinel, Florian and Shwartz, Larisa and Grabarnik, Genady},
  booktitle={2012 IEEE Network Operations and Management Symposium},
  pages={34--42},
  year={2012},
  url          = {https://doi.org/10.1109/NOMS.2012.6211880},
  doi          = {10.1109/NOMS.2012.6211880},
  organization={IEEE}
}

@inproceedings{lin2018collaborative,
  title={Collaborative alert ranking for anomaly detection},
  author={Lin, Ying and Chen, Zhengzhang and Cao, Cheng and Tang, Lu-An and Zhang, Kai and Cheng, Wei and Li, Zhichun},
  booktitle={CIKM 2018},
  url          = {https://doi.org/10.1145/3269206.3272013},
  doi          = {10.1145/3269206.3272013},
  pages={1987--1995},
  year={2018}
}

@inproceedings{aminanto2019automated,
  title={Automated threat-alert screening for battling alert fatigue with temporal isolation forest},
  author={Aminanto, Muhamad Erza and Zhu, Lei and Ban, Tao and Isawa, Ryoichi and Takahashi, Takeshi and Inoue, Daisuke},
  booktitle={PST 2019},
  url          = {https://doi.org/10.1109/PST47121.2019.8949029},
  doi          = {10.1109/PST47121.2019.8949029},
  pages={1--3},
  year={2019},
  organization={IEEE}
}

@article{almeida2019word,
  author       = {Felipe Almeida and
                  Geraldo Xex{\'{e}}o},
  title        = {Word Embeddings: {A} Survey},
  journal      = {CoRR},
  volume       = {abs/1901.09069},
  year         = {2019},
  url          = {http://arxiv.org/abs/1901.09069},
}

@inproceedings{transformer,
  author       = {Ashish Vaswani and
                  Noam Shazeer and
                  Niki Parmar and
                  Jakob Uszkoreit and
                  Llion Jones and
                  Aidan N. Gomez and
                  Lukasz Kaiser and
                  Illia Polosukhin},
  title        = {Attention is All you Need},
  booktitle    = {Neurips 2017},
  pages        = {5998--6008},
  url          = {https://proceedings.neurips.cc/paper/2017/hash/3f5ee243547dee91fbd053c1c4a845aa-Abstract.html},
  year         = {2017},
}

@inproceedings{liu2008isolation,
  title={Isolation forest},
  author={Liu, Fei Tony and Ting, Kai Ming and Zhou, Zhi-Hua},
  booktitle={ICDM 2008},
  pages={413--422},
  year={2008},
  url          = {https://doi.org/10.1109/ICDM.2008.17},
  doi          = {10.1109/ICDM.2008.17},
  organization={IEEE}
}

@article{campello2015hierarchical,
  title={Hierarchical density estimates for data clustering, visualization, and outlier detection},
  author={Campello, Ricardo JGB and Moulavi, Davoud and Zimek, Arthur and Sander, J{\"o}rg},
  journal={ACM Transactions on Knowledge Discovery from Data (TKDD)},
  volume={10},
  number={1},
  pages={1--51},
  year={2015},
  url          = {https://doi.org/10.1145/2733381},
  doi          = {10.1145/2733381},
  publisher={ACM}
}

@inproceedings{paszke2019pytorch,
  author       = {Adam Paszke and
                  Sam Gross and
                  Francisco Massa and
                  Adam Lerer and
                  James Bradbury and
                  Gregory Chanan and
                  Trevor Killeen and
                  Zeming Lin and
                  Natalia Gimelshein and
                  Luca Antiga and
                  Alban Desmaison and
                  Andreas K{\"{o}}pf and
                  Edward Z. Yang and
                  Zachary DeVito and
                  Martin Raison and
                  Alykhan Tejani and
                  Sasank Chilamkurthy and
                  Benoit Steiner and
                  Lu Fang and
                  Junjie Bai and
                  Soumith Chintala},
  title        = {PyTorch: An Imperative Style, High-Performance Deep Learning Library},
  booktitle    = {NeurIPS 2019},
  pages        = {8024--8035},
  url          = {https://proceedings.neurips.cc/paper/2019/hash/bdbca288fee7f92f2bfa9f7012727740-Abstract.html},
  year         = {2019}
}

@inproceedings{siffer2017anomaly,
  title={Anomaly detection in streams with extreme value theory},
  author={Siffer, Alban and Fouque, Pierre-Alain and Termier, Alexandre and Largouet, Christine},
  booktitle={KDD 2017},
  pages={1067--1075},
  url          = {https://doi.org/10.1145/3097983.3098144},
  doi          = {10.1145/3097983.3098144},
  year={2017}
}

@misc{prometheus,
  title        = {The Prometheus monitoring system and time series database},
  howpublished = {\url{https://github.com/prometheus/prometheus}},
  year = 2025,
}

@misc{alertmanager,
  title        = {Alertmanager: An open-source monitoring system},
  howpublished = {\url{https://github.com/prometheus/alertmanager}},
  year = 2025,
}

@misc{grafana,
  title        = {Compose and scale your observability with one, some, or all of the Grafana Stack pieces},
  howpublished = {\url{https://grafana.com/}},
  year = 2025,
}

@misc{PromQL,
  title        = {PromQL (Prometheus Query Language)},
  howpublished = {\url{https://prometheus.io/docs/prometheus/latest/querying/basics/}},
  year = 2025
}

@misc{LogsQL,
  title        = {LogsQL (Logs Query Language)},
  howpublished = {\url{https://docs.victoriametrics.com/victorialogs/logsql/}},
  year = 2025
}

@misc{DeepseekR1,
  title        = {DeepSeek R1},
  howpublished = {\url{https://github.com/deepseek-ai/DeepSeek-R1}},
  year = 2025
}

@misc{DeepseekV3,
  title        = {DeepSeek V3},
  howpublished = {\url{https://github.com/deepseek-ai/DeepSeek-V3}},
  year = 2025
}

@misc{K2view,
  title        = {GenAI Adoption 2024: The Challenge with Enterprise Data},
  howpublished = {\url{https://www.k2view.com/genai-adoption-survey/}},
  year = 2025
}

@misc{azurerule,
  title        = {Azure Rule Management},
  howpublished = {\url{https://learn.microsoft.com/en-us/azure/azure-monitor/alerts/alerts-manage-alert-rules}},
  year = 2025
}

@misc{googlerule,
  title        = {Google Cloud Rule Management},
  howpublished = {\url{https://cloud.google.com/distributed-cloud/hosted/docs/latest/appliance/admin/create-alert-rules}},
  year = 2025
}

@inproceedings{COT,
  author       = {Jason Wei and
                  Xuezhi Wang and
                  Dale Schuurmans and
                  Maarten Bosma and
                  Brian Ichter and
                  Fei Xia and
                  Ed H. Chi and
                  Quoc V. Le and
                  Denny Zhou},
  title        = {Chain-of-Thought Prompting Elicits Reasoning in Large Language Models},
  booktitle    = {NeurIPS 2022},
  year         = {2022},
  url          = {http://papers.nips.cc/paper\_files/paper/2022/hash/9d5609613524ecf4f15af0f7b31abca4-Abstract-Conference.html},
}

@misc{gu2025argosagentictimeseriesanomaly,
      title={Argos: Agentic Time-Series Anomaly Detection with Autonomous Rule Generation via Large Language Models}, 
      author={Yile Gu and Yifan Xiong and Jonathan Mace and Yuting Jiang and Yigong Hu and Baris Kasikci and Peng Cheng},
      year={2025},
      eprint={2501.14170},
      archivePrefix={arXiv},
      primaryClass={cs.LG},
      url={https://arxiv.org/abs/2501.14170}, 
}

@inproceedings{MicroSketch,
  author       = {Yufeng Li and
                  Guangba Yu and
                  Pengfei Chen and
                  Chuanfu Zhang and
                  Zibin Zheng},
  title        = {MicroSketch: Lightweight and Adaptive Sketch Based Performance Issue
                  Detection and Localization in Microservice Systems},
  booktitle    = {{ICSOC} 2022},
  series       = {Lecture Notes in Computer Science},
  volume       = {13740},
  pages        = {219--236},
  publisher    = {Springer},
  year         = {2022},
  doi          = {10.1007/978-3-031-20984-0\_15},
}

@inproceedings{tang2015line,
  title={Line: Large-scale information network embedding},
  author={Tang, Jian and Qu, Meng and Wang, Mingzhe and Zhang, Ming and Yan, Jun and Mei, Qiaozhu},
  booktitle={WWW 2015},
  pages={1067--1077},
  url          = {https://doi.org/10.1145/2736277.2741093},
  doi          = {10.1145/2736277.2741093},
  year={2015}
}

@inproceedings{moritz2018ray,
  title={Ray: A distributed framework for emerging AI applications},
  author={Moritz, Philipp and Nishihara, Robert and Wang, Stephanie and Tumanov, Alexey and Liaw, Richard and Liang, Eric and Elibol, Melih and Yang, Zongheng and Paul, William and Jordan, Michael I and others},
  booktitle={OSDI 2018},
  pages={561--577},
  url          = {https://www.usenix.org/conference/osdi18/presentation/nishihara},
  year={2018}
}

@inproceedings{kingma2014adam,
  title={Adam: A method for stochastic optimization},
  author={Kingma, Diederik P and Ba, Jimmy},
  booktitle    = {{ICLR} 2015},
  year         = {2015},
  url          = {http://arxiv.org/abs/1412.6980},
}

@article{rag,
  author       = {Yunfan Gao and
                  Yun Xiong and
                  Xinyu Gao and
                  Kangxiang Jia and
                  Jinliu Pan and
                  Yuxi Bi and
                  Yi Dai and
                  Jiawei Sun and
                  Qianyu Guo and
                  Meng Wang and
                  Haofen Wang},
  title        = {Retrieval-Augmented Generation for Large Language Models: {A} Survey},
  journal      = {CoRR},
  volume       = {abs/2312.10997},
  year         = {2023},
  url          = {https://doi.org/10.48550/arXiv.2312.10997},
  doi          = {10.48550/ARXIV.2312.10997}
}

@inproceedings{li2022incident,
  title={Going through the Life Cycle of Faults in Clouds: Guidelines on Fault Handling},
  author={Li, Xiaoyun and Yu, Guangba and Chen, Pengfei and Chen, Hongyang and Chen, Zhekang},
  booktitle={ISSRE 2022},
  pages={121--132},
  year={2022},
  publisher={IEEE},
  doi          = {10.1109/ISSRE55969.2022.00022},
  url          = {https://doi.org/10.1109/ISSRE55969.2022.00022},
}

@inproceedings{socc2022incident,
  author       = {Supriyo Ghosh and
                  Manish Shetty and
                  Chetan Bansal and
                  Suman Nath},
  title        = {How to fight production incidents?: an empirical study on a large-scale
                  cloud service},
  booktitle    = {SoCC 2022},
  pages        = {126--141},
  publisher    = {{ACM}},
  year         = {2022},
  url          = {https://doi.org/10.1145/3542929.3563482},
  doi          = {10.1145/3542929.3563482},
}

@inproceedings{nezha2023fse,
  author       = {Guangba Yu and
                  Pengfei Chen and
                  Yufeng Li and
                  Hongyang Chen and
                  Xiaoyun Li and
                  Zibin Zheng},
  title        = {Nezha: Interpretable Fine-Grained Root Causes Analysis for Microservices on Multi-modal Observability Data},
  booktitle    = {{ESEC/FSE} 2023},
  pages        = {553--565},
  publisher    = {{ACM}},
  year         = {2023},
  url          = {https://doi.org/10.1145/3611643.3616249},
  doi          = {10.1145/3611643.3616249},
}

@article{changerca2024fse,
  author       = {Guangba Yu and
                  Pengfei Chen and
                  Zilong He and
                  Qiuyu Yan and
                  Yu Luo and
                  Fangyuan Li and
                  Zibin Zheng},
  title        = {ChangeRCA: Finding Root Causes from Software Changes in Large Online
                  Systems},
  journal      = {Proc. {ACM} Softw. Eng.},
  volume       = {1},
  number       = {{FSE}},
  pages        = {24--46},
  year         = {2024},
  url          = {https://doi.org/10.1145/3643728},
  doi          = {10.1145/3643728},
}

@inproceedings{COLA2024ICSE,
  author       = {Jinxi Kuang and
                  Jinyang Liu and
                  Junjie Huang and
                  Renyi Zhong and
                  Jiazhen Gu and
                  Lan Yu and
                  Rui Tan and
                  Zengyin Yang and
                  Michael R. Lyu},
  title        = {Knowledge-aware Alert Aggregation in Large-scale Cloud Systems: a
                  Hybrid Approach},
  booktitle    = {{ICSE-SEIP} 2024},
  pages        = {369--380},
  publisher    = {{ACM}},
  year         = {2024},
  url          = {https://doi.org/10.1145/3639477.3639745},
  doi          = {10.1145/3639477.3639745}
}

@inproceedings{logreducer,
  author       = {Guangba Yu and
                  Pengfei Chen and
                  Pairui Li and
                  Tianjun Weng and
                  Haibing Zheng and
                  Yuetang Deng and
                  Zibin Zheng},
  title        = {LogReducer: Identify and Reduce Log Hotspots in Kernel on the Fly},
  booktitle    = {{ICSE} 2023},
  pages        = {1763--1775},
  publisher    = {{IEEE}},
  year         = {2023},
  url          = {https://doi.org/10.1109/ICSE48619.2023.00151},
}

@misc{preserve,
      title={Hierarchical Prediction-based Management for LMaaS Systems}, 
      author={Zhihan Jiang and Yujie Huang and Guangba Yu and Junjie Huang and Jiazhen Gu and Michael R. Lyu},
      year={2025},
      eprint={2504.03702},
      archivePrefix={arXiv},
      url={https://arxiv.org/abs/2504.03702}, 
}

@ARTICLE{yu2020microscaler2,  
  author={Yu, Guangba and Chen, Pengfei and Zheng, Zibin},  
  journal={IEEE TCC},   
  title={Microscaler: Cost-effective Scaling for Microservice Applications in the Cloud with an Online Learning Approach},   
  publisher = {{IEEE}},
  year={2022},
  volume={10},
  number={2},
  pages={1100-1116},
  doi={10.1109/TCC.2020.2985352},
 url = {https://doi.org/10.1109/TCC.2020.2985352}
}

@inproceedings{yu2021microrank,
  author = {Yu, Guangba and Chen, Pengfei and Chen, Hongyang and Guan, Zijie and Huang, Zicheng and Jing, Linxiao and Weng, Tianjun and Sun, Xinmeng and Li, Xiaoyun},
  title = {MicroRank: End-to-End Latency Issue Localization with Extended Spectrum Analysis in Microservice Environments},
  year = {2021},
  publisher = {ACM},
  booktitle = {WWW 2021},
  pages = {3087–3098},
  numpages = {12},
  doi      = {10.1145/3442381.3449905},
}

@inproceedings{He2022GIED,
author = {He, Zilong and Chen, Pengfei and Luo, Yu and Yan, Qiuyu and Chen, Hongyang and Yu, Guangba and Li, Fangyuan},
title = {Graph based Incident Extraction and Diagnosis in Large-Scale Online Systems},
year = {2023},
publisher = {ACM},
url = {https://doi.org/10.1145/3551349.3556904},
doi = {10.1145/3551349.3556904},
booktitle = {ASE 2022},
articleno = {48},
}

\end{document}